\documentclass{aa}  
\usepackage{graphicx}
\usepackage{txfonts}
\usepackage{natbib}
\begin{document} 

   \title{Atmospheric NLTE-models for the spectroscopic analysis of blue
stars with winds}
   \subtitle{IV. Porosity in physical and velocity space}

   \author{J.O. Sundqvist\inst{1}\and
     J. Puls\inst{2}}

   \institute{KU Leuven, Instituut voor Sterrenkunde, Celestijnenlaan 200D, 3001 Leuven, 
   Belgium\\ \email{jon.sundqvist@kuleuven.be}\and LMU M\"unchen, Universit\"atssternwarte, Scheinerstr. 1,
     81679 M\"unchen, Germany}
                                 
   \date{Received 2018-03-09; accepted 2018-05-26}

% \abstract{}{}{}{}{} 
% 5 {} token are mandatory
 
  \abstract
  % context heading (optional)
   {Clumping in the radiation-driven winds of hot, massive stars severly affects the derivation of 
   synthetic observables across the electromagnetic spectrum.} %leave it empty if necessary 
  % aims heading (mandatory) 
   {We implement a formalism for treating wind clumping -- focusing in particular on the light-leakage effects associated with a medium   
   that is porous in physical and velocity space -- into the global (photosphere+wind) NLTE 
   model atmosphere and spectrum synthesis code {\sc fastwind}.}
   %and to investigate its effects on the theory and
   %diagnostics of line-driven winds from hot, massive stars.}
  % methods heading (mandatory) 
   {The basic method presented here assumes a stochastic, two-component wind consisting of a 
   mixture of optically thick and thin clumps embedded in a rarefied inter-clump medium. We have accounted 
   fully for the reductions 
   in opacity associated with porosity in physical and velocity-space (the latter due to Doppler shifts 
   in an accelerating medium), as well as for the well-known effect that opacities depending 
   on $\langle \rho^2 \rangle$ are higher in clumpy winds than in smooth ones   
   of equal mass-loss rate. By formulating our method in terms of suitable mean and 
   effective opacities for the clumpy wind, we are able to compute atmospheric models 
   with the same speed  ($\sim15$ minutes on a modern laptop or desktop) as 
   in previous generations of {\sc fastwind}.}
  % results heading (mandatory) 
   %The reduction in effective
   %line-opacity in an accelerating two-compoent stellar wind stems
   %from porous channels in velocity space through which light escapes
   %without ever interacting with the dense clumps. The conceptual
   %difference between this velocity-porosity (=vorosity) and the
   %spatial porosity responsible for continuum opacity redcution is
   %reflected in the calculation of the clump optical depth, which for
   %continuum depnds on the mean-free-path between clumps (a.k.a. the
   %porosity length) but for lines on the velocity factor $f_{\rm
   %  vel}$, defined here as the velocity width of clumps divided by
   %their mean velocity separation.  
   {After verifying important analytic limits (smooth, optically thin, 
   completely optically thick), we present some first, generic results of the new 
   models. These include: i) Confirming earlier results that velocity-space porosity 
   is critical for analysis of UV wind resonance lines in O-stars; ii) for the 
   optical H$\alpha$ line, we show that optically thick clumping effects 
   are small for O-stars, but potentially very important for late 
   B and A-supergiants; iii) in agreement with previous work, we show 
   that spatial porosity is a marginal effect for absorption of high-energy X-rays in 
   O-stars, as long as the mean-free path between clumps are kept at realistic values 
   $\la R_\ast$; iv) whereas radio absorption in O-stars shows strong spatial porosity effects in 
   near photospheric layers, it is negligible at their typical radio-photosphere radii 
   $\sim 100 R_\ast$; v) regarding the wind ionization balance, a general trend
   is that increased rates of recombination in simulations with optically thin clumps lead to overall lower 
   degrees of ionization than in corresponding smooth models, but that this effect 
   now is counteracted by the increased levels of light-leakage associated with 
   porosity in physical and velocity space (i.e., by an increase of ionization rates). We conclude by discussing future work and 
   some planned applications for this new generation of {\sc fastwind} models.}
   %conclusions heading (optional)
{}
% leave it empty
   \maketitle
%
%________________________________________________________________

\section{Introduction}
\label{intro}

Information about fundamental parameters of stars -- like their mass, luminosity,
surface temperature and chemical composition -- comes primarily from
matching observations to synthetic spectra computed using
models of stellar atmospheres. For massive stars with hot surfaces,
scattering and absorption in spectral lines further transfer momentum from the
star's intense radiation field to the plasma, and so provides a 
force that overcomes gravity and drives a wind-outflow directly from the 
stellar surface \citep[][]{Castor75}. These starlight-powered winds are very strong 
and fast, and can dramatically affect the star's atmospheric structure
\citep[review by][]{Puls08} as well as the evolution of 
its mass and luminosity, chemical surface abundances,
rotational velocity, and nuclear burning life-times \citep[review by][]{Smith14}. 

Atmospheric models of such hot, massive stars 
must thus generally be constructed using a unified, or global, 
approach, wherein the basic structural equations for the quasi-static 
photosphere and the outflowing stellar wind 
are solved simultaneously \citep{Gabler89}. In addition, 
the expanding atmospheres of these stars are characterized by 
their large departures from local thermodynamic equilibrium 
(LTE), meaning the full number density rate equations (typically reduced to statistical equilibrium, and often 
simply called non-LTE or NLTE) must be solved to obtain the 
atmospheric radiation field and the excitation and ionization balance. As such, 
quite intricate numerical solution techniques are normally required to compute synthetic 
observables, like spectral lines and energy 
distributions, for these objects \citep[for details, see book by][]{Hubeny14}. 

Over the past decades, much effort has been devoted toward  
constructing such global NLTE, steady-state model atmospheres of hot stars with 
winds; several numerical 
computer codes meanwhile exist on the market,  
for example {\sc cmfgen} \citep{Hillier98}, {\sc PoWR} 
(\citealt{Grafener02, Sander15}), {\sc phoenix} \citep{Hauschildt92}, 
{\sc WM-Basic} \citep{Pauldrach01}, 
and the subject of this paper, {\sc fastwind} 
\citep{Santolaya97, Puls05, Rivero11, Carneiro16}. {\sc fastwind} 
is routinely applied for both photospheric and wind analyses of
hot stars, and used for detailed studies of individual objects as well
as in large spectroscopic surveys (like within the recent {\sc
  vlt-flames} survey of massive stars in the Tarantula giant
star-forming region in the Large Magellanic Cloud, \citealt{Evans11}). 

A critical component of all these codes regards their practical 
treatment of the stellar wind; traditionally this has been to assume a  
parametrized steady-state and smooth outflow, without any clumps or 
shocks. However, it has been known for quite many years now, that 
these line-radiation driven winds are in fact inhomogeneous
and highly structured on small spatial scales \citep[see overviews
  in][]{Puls08, Hamann08, Sundqvist12c, Puls15}. Such wind clumping
arises naturally from the strong line-deshadowing instability, the
LDI, a fundamental and inherent property of line driving \citep[e.g.,][]{Owocki84,
  Owocki85}. Radiation-hydrodynamic, time-dependent 
wind models \citep{Owocki88, Feldmeier97, Owocki99, Dessart03, Sundqvist13, Sundqvist15,
Sundqvist18} 
following the non-linear evolution of this LDI show a 
characteristic two-component-like structure consisting of 
spatially small and dense clumps separated by large regions of 
very rarified material, accompanied by strong thermal shocks and 
a highly non-monotonic velocity field. Such clumpy winds then 
affect both the atmospheric structure and the radiative transfer needed 
to derive synthetic observables; as just one example of this,  
neglecting clumping typically leads to observationally inferred mass-loss rates
that might differ by more than an order of magnitude for the same star,
depending on which spectral diagnostic is used to estimate this mass
loss \citep{Fullerton06}.

Global model atmospheres nowadays normally account for 
such wind inhomogeneities by simply assuming a two-component 
medium consisting of overdense, optically thin clumps of a 
certain volume filling factor, following a smooth, parametrized velocity law, and 
an inter-clump medium that is effectively void 
\citep[e.g.,][]{Hillier91, Puls06}. However, if clumps become optically 
thick, it leads to an additional leakage of light -- 
not accounted for in the filling factor approach -- through porous 
channels in between the clumps. Such porosity can occur either 
spatially \citep[e.g.,][]{Feldmeier03, Owocki04, Sundqvist12}, or for 
spectral lines in \textit{velocity-space} due to Doppler shifts in the 
rapidly accelerating wind (sometimes thus called velocity-porosity, or
``vorosity'', \citealt{Owocki08}). Regarding spatial porosity, several
studies over the past years have focused on examining potential 
effects on the bound-free absorption of X-ray photons by the 
bulk wind \citep[e.g.,][]{Oskinova06, Owocki06, Sundqvist12, 
Leutenegger13, Herve13}. Regarding velocity-space porosity, similar 
studies \citep{Oskinova07, Hillier08, Sundqvist10, Sundqvist11, 
  Surlan12, Surlan13, Sundqvist14} have shown that 
  clumps indeed very easily become optically thick
in especially the strong UV wind-lines of hot stars (the so-called P-Cygni
lines), and that the associated additional 
leakage of line-photons leads to weaker line profiles than predicted by smooth or
volume filling factor models.\footnote{A note on terminology: Some authors 
collect the above effects of optically thick clumps under the umbrella-term 
``macroclumping", and similarly name the case of only 
optically thin clumps ``microclumping". This essentially derives from 
the ``microturbulence"  and ``macroturbulence" terminology traditionally used in 
spectroscopy of stellar photospheres. However, due to the quite different 
properties of line and continuum clumping-effects in accelerating media, 
as well as the risk of confusing  optical depths with spatial scales within the
micro- and macro-clumping terminology, this paper simply uses the denotation 
optically thin and thick clumping throughout, and describes effects of the latter 
using the more physical terms porosity in spatial and/or velocity-space.}  

But constructing realistic, multi-dimensional \textit{ab-initio}
radiation-hydrodynamic wind simulations that account naturally for
(time-dependent) spatial and velocity-field porosity is an extremely challenging and
time-consuming task \citep{Sundqvist18}. Thus there 
has also been a big need for developing
simplified, parameterized models that can be more routinely applied to
diagnostic work on samples of hot stars with winds. 
Building on their prior studies \citep{Sundqvist10, Sundqvist11, 
Sundqvist12}, \citet{Sundqvist14} (hereafter SPO14) developed and 
benchmarked such a method, using effective quantities 
to simulate the reduction in opacity associated with optically thick 
clumps. In contrast to some other models mentioned above, this `effective
opacity' approach has the great advantage that it can be quite readily
implemented into the already existing (time-independent) 
global NLTE atmosphere models discussed above. 

This paper incorporates the SPO14 formalism 
into the {\sc fastwind} computer code, and presents 
some first results. In \S 2 we briefly review the basic 
physics of {\sc fastwind}, and \S 3 describes our 
methodology and practical implementation of 
wind clumping into the code. \S 4 then presents some first 
results, and in \S 5 we summarize and outline future work.  

\section{Basic physics of the global model atmosphere code {\sc fastwind}}
\label{fastwind} 

The versatile and fast (computing time $\sim$15 min on a
modern desktop/laptop) global model atmosphere computer code {\sc fastwind} 
solves the NLTE number-density rate equations within a
spherically extended envelope containing both the stellar photosphere
and the supersonic average wind, and including the effects from
millions of metal spectral lines on the atmospheric structure. The present 
version is designed for analyzing stars of spectral types OBA,  
with winds that are not significantly optically thick in the 
optical continuum\footnote{though an update is currently under way to treat 
also stars with dense winds, like Wolf-Rayet stars (Sundqvist et al., 
in prep.).}. 

%\subsection{Basic physics}

\subsection{Momentum and energy balance} 

As described 
in \citet{Santolaya97}, in the deep atmosphere 
{\sc fastwind} neglects the advection term in the momentum 
equation and instead solves the equation of hydrostatic 
equilibrium in spherical symmetry:  
\begin{equation} 
	\varv \frac{d \varv}{dr} \approx 0 = - \frac{dP}{dr}(r)\frac{1}{\rho(r)} - g_\ast \left( \frac{R_\ast}{r} \right)^2 + g_{\rm rad}(r),  
\end{equation}   
where $g_\ast$ and $R_\ast$ are the (input parameters)
surface gravity and radius, $\rho$ is the mass density, 
$dP/dr$ the gas pressure gradient, and $g_{\rm rad}$ 
the radiative acceleration in the photosphere. A novel 
feature of the basic methodology behind {\sc fastwind} 
is the calculation of the flux-weighted opacities, needed 
above to compute $g_{\rm rad}$, which 
%(for not too optically thick 
%winds) 
in these deep atmospheric layers can be well 
approximated with a Kramer's like opacity-formula
(see \citealt{Santolaya97}). 

The photospheric structure is 
smoothly connected to the wind outflow at a pre-specified 
velocity, typically set to $\approx$ 10\,\% of the isothermal 
sonic speed $a$ for the assumed (input parameter) stellar effective temperature,
$\varv_{\rm trans} \approx 0.1 a(T_{\rm eff})$. The version 
of {\sc fastwind} presented here further 
\textit{pre-specifies}\footnote{though an 
update of also this is under way, in which the dynamical 
equations are actually solved in the wind (Sundqvist et al., in prep.).}
the wind structure by adopting a '$\beta$' velocity law and 
a wind mass-loss rate $\dot{M}$, i.e.,  
\begin{equation} 
	\varv(r) = \varv_\infty (1-b/r)^{\beta}, 
\end{equation} 
\begin{equation} 
	\rho(r) = \dot{M}/(4 \pi \varv(r) r^2), 
	\label{Eq:mdot}		
\end{equation}
where $\varv_\infty$ is the assumed terminal wind speed, and 
$b$ is obtained from the calculated radius at the assumed wind 
boundary $\varv_{\rm trans}$;  the velocity field of the quasi-static photosphere 
then follows directly from mass conservation using the computed 
photospheric density. 

As described in \citet{Puls05}, a consistent energy balance (temperature structure)  
is computed in parallel, using a flux-correction method in the 
lower atmosphere and the thermal balance of free electrons in the 
outer. As usual, the system of equations is closed by the ideal  
gas law, $P(r) = \rho(r) a^2(r) = \rho(r) k_{\rm B} T(r)/(\mu(r) m_{\rm H})$ for mean 
molecular weight $\mu(r)$ and with Boltzmann's constant 
$k_{\rm B}$ and hydrogen atomic mass $m_{\rm H}$.     

We note further that when introducing wind clumping as 
described in \S3, $\rho$ and $\varv$ 
in the equations above essentially are the 
\textit{mean} density and velocity, respectively. 

\subsection{Detailed model atoms and background elements} 

To enable shorter computational times, {\sc fastwind} separates 
between 'explicit' elements considered in great detail and 
'background' elements treated in a somewhat more approximate 
way. Explicit elements use detailed model atoms and 
co-moving frame radiation transport for all line transitions, whereas 
the background elements use parameterized ionization cross-sections
(see \citealt{Puls05}) and co-moving frame transfer only for 
the most important, strong lines, whereas the 
rest of the lines are calculated using the Sobolev (1960)
approximation.\footnote{Also here is a code-update 
under way, which will compute the full radiation field 
by means of co-moving frame transport \citep{Puls17}.} 
Model atoms for the explicit elements are 
provided by the user, and are typically those 
used for various diagnostic spectroscopic purposes 
(H, He, N, P, etc; e.g., most recently a new carbon model atom 
has been implemented, see \citealt{Carneiro17}). By contrast, the background elements 
are needed "only" for a consistent description of line blocking/blanketing (see below)
and their data are pre-specified by our atomic data base \citep{Pauldrach01}; this background 
then includes (almost) all elements up to Zn not treated as explicit ones in the 
particular calculations (for a more detailed 
description of the basic philosophy behind explicit and 
background elements, see \citealt{Puls05} and \citealt{Rivero11}).   
As is customary, also the metallicity and chemical composition of the 
atmosphere (including the helium abundance) are provided by the user, where 
we adopt solar abundances from \cite{asplund09} as default. 

\subsection{Line blocking and blanketing} 

As already mentioned, 
a key point of the philosophy behind {\sc fastwind} is the \textit{speed} 
of the program, which allows us to perform calculations much 
faster than comparable NLTE codes available on the market. In
addition to the explicit/background element-approach described above, 
this computational speed is to a large extent due to the novel way of 
treating the opacity-effects of millions of weak spectral lines 
developed by \citet{Puls05}. In short, this method uses 
the way described above for solving the NLTE equations for the background elements, 
and then introduces a simple statistical approach for calculating 
the line opacities and emissivities (based on suitable averages) 
that are used within the radiative transfer to account for line 
blocking/blanketing \citep[details given in][]{Puls05}. 

\subsection{X-ray emission from embedded wind-shocks} 

Massive, hot stars are ubiquitous sources of high-energy X-ray emission, typically 
on order $L_{\rm x}/L_{\rm Bol} \sim 10^{-7}$ \citep[e.g.,][]{Rauw15b}. For 
putatively single, non-magnetic 
massive stars these X-rays are due to strong shocks embedded in 
the ambient stellar wind, and associated with 
the same fundamental instability of line-driven stellar winds 
that causes wind clumping (the LDI, see introduction). Recently, a 
module has been incorporated into {\sc fastwind}
that computes the ionization and absorption effects on the bulk wind
stemming from such X-ray emission \citep{Carneiro16}. In summary, in this method 
a (very) small fraction of the stellar wind is assumed to emit high-energy photons 
by means of thermal shocks, whereafter both the direct and Auger ionization effects 
arising from this are calculated, as well as the absorption (via valence and 
K-shell electrons) of the emitted high-energy photons by 
the ambient wind (for further details, see \citealt{Carneiro16}).\\

\noindent Having summarized essential features and physical assumptions, the 
next section now describes in detail our 
implementation of wind clumping into {\sc fastwind}. 

\section{Treatment of wind clumping}
\label{clumping} 

Our treatment of wind-clumping follows the formalism developed by 
SPO14. As shown in detail below, this methodology enables a simple and fast 
implementation, in both the NLTE rate equations and the computation
of synthetic spectra, which accounts for the effects of porosity in physical 
and velocity space while preserving the (previously considered) limits 
of smooth winds and winds consisting of only optically thin 
clumps. Two key simplifying assumptions of our model are:       

\paragraph{One-component average ionization description.} Numerical NLTE computations 
like those performed by {\sc fastwind} and similar computer codes 
are very challenging, and require much computational time. To avoid 
a full multi-component NLTE calculation of the 
structured wind\footnote{which would be \textit{very} complex and time-demanding, and at the 
moment almost impossible for practical applications.}, we assume here that the 
wind excitation and ionization balance can be approximated from considering a 
suitable, single `effective' opacity of a two-component medium, accounting for  
contributions both from the dense clumps and the rarified medium in between 
them. As shown by SPO14 (see also \citealt{Pomraning91}), using such 
`effective' quantities then allows us to simply re-scale the opacities of 
non-structured simulations, and so calculate new models that account for 
clumps of \textit{arbitrary} optical depths without any essential loss of 
computation-time. While neglecting potential effects of a multi-component 
ionization balance \citep[see][]{Zsargo08}, this approach is thus very advantageous 
regarding, e.g., practical applications such as spectroscopy of larger 
stellar samples. 

\paragraph{Absorption and emission.} The formalism below is developed 
for absorption processes. An important assumption of our model 
is that corresponding emission coefficients are scaled analogously, 
using the same expressions for the clump optical depths (see next
section). As shown by \citet{Pomraning91}, this is indeed exact 
in the limit of only thermal absorption in a two-component model 
where both components have the same temperature. Although 
this procedure is formally not exact for scattering (see also discussion 
in \citealt{Pomraning91}, their Ch. 4), we here scale \textit{all} emission 
processes in the same way, guided by the results of our 
previous multi-dimensional, Monte-Carlo line-scattering 
simulations \citep{Sundqvist10, Sundqvist11, Sundqvist14}.\\ 
    
\noindent Having outlined above these two important physical assumptions, 
the below describes in detail our model. Limiting cases (particularly regarding 
an effectively void interclump medium) are provided in \S4.      
    
\subsection{Mean and effective opacities} 

For a stochastic two-component medium consisting of dense clumps 
(hereafter subscripts 'cl') occupying a filling factor 
$f_{\rm vol}$ of the total volume, and a rarified 'inter-clump' (subscript 'ic') 
medium filling in the space in between clumps, the mean mass density 
is 
\begin{equation} 
	\langle \rho \rangle = f_{\rm vol} \rho_{\rm cl} + (1-f_{\rm vol}) \rho_{\rm ic} = \rho_{\rm sm} % = \frac{\dot{M}}{4 \pi r^2 \varv},  
	\label{Eq:rhom}
\end{equation} 
where the last equality assumes mass-conservation with respect 
to a smooth medium of density $\rho_{\rm sm}$. 
For notational 
simplicity, all expressions throughout this section
suppress dependencies on the local radius.  
The mean density is 
related to the mean-square density via the so-called clumping factor  
\begin{equation}
	f_{\rm cl} \equiv \frac{\langle \rho^2 \rangle}{\langle \rho \rangle^2} = 
	\frac{f_{\rm vol} \rho_{\rm cl}^2 + (1-f_{\rm vol}) \rho_{\rm ic}^2}
	{(f_{\rm vol} \rho_{\rm cl} + (1-f_{\rm vol}) \rho_{\rm ic})^2}
	\ge 1.   
	\label{Eq:fcl}
\end{equation} 
To avoid a full two-component NLTE computation, this paper introduces  
a formalism that preserves known limiting cases (see below) for the 
\textit{mean opacity} per unit length, 
$\langle \chi \rangle = \kappa \langle \rho \rangle$ with 
mass absorption coefficient $\kappa$,  
for processes where $\langle \chi \rangle \sim \langle \rho \rangle$ 
and $\langle \chi \rangle \sim \langle \rho^2 \rangle$.  

The mean opacity is obtained by first computing 
the NLTE occupation numbers $n$
within a ``fiducial'' clump of density $\rho_f = \langle \rho \rangle f_{\rm
  cl} = \rho_{\rm sm} f_{\rm cl}$, and then reducing the fiducial clump
opacity $\chi_f \equiv \sigma n_f$ by a clumping factor $f_{\rm cl}$, 
\begin{equation} 
  \langle \chi \rangle = \chi_f/f_{\rm cl}, %\sigma \  n_f/f_{\rm cl} = \kappa \langle \rho \rangle,    
  \label{Eq:chi_mean} 
\end{equation}   
where $\sigma$ and $n_{\rm f}$ above are the atomic cross-section and the number 
density of the fiducial
clump, respectively, and where the underlying "smooth" 
scaling density $\rho_{\rm sm} = \dot{M}/(4 \pi \varv r^2)$ is set by the wind mass-loss rate 
(eqn. \ref{Eq:mdot}). We recall here again that this "average medium" treatment implicitly assumes that the 
(gas and radiation) temperatures of clump and inter-clump media 
are similar (see also above). 
%where the second equality introduces the \textit{mass} absorption 
%coefficient $\kappa = \sigma n_f / \rho_f$. 
Formulated this way, eqn.~\ref{Eq:chi_mean} preserves the well-known results 
that $\langle \chi \rangle$ is unaffected by clumping if 
$\chi = \sigma n \sim \rho$ and enhanced (as compared to an unclumped medium) 
by a factor $f_{\rm cl}$ if $\chi = \sigma n \sim \rho^2$, as can be shown by 
a simple calculation (see also Eqs. 25/26).
%$n \sim \rho$ and enhanced by factor of $f_{\rm cl}$ if $n \sim \rho^2$.

Using this mean opacity, we then write the \textit{effective} 
opacity of the two component medium as (SPO14) 
\begin{equation} 
  \chi_{\rm eff} = \langle \chi \rangle \frac{1 + \tau_{\rm cl} f_{\rm ic}}{1+\tau_{\rm cl}}, 
  %\equiv  \langle \chi \rangle \, f_{\rm red}(\tau_{\rm cl}, f_{\rm ic}),
  \label{Eq:chi_eff}
\end{equation}   
where $\tau_{\rm cl}$ is the clump optical depth and 
\begin{equation} 
	f_{\rm ic} \equiv \rho_{\rm ic}/\langle \rho \rangle
 	\label{Eq:fic} 
\end{equation}
sets the density of the rarefied medium in between the clumps. We note 
that Eq.~\ref{Eq:fic} gives the inter-clump density in terms of the 
mean density; the density 
contrast between clumps and inter-clump medium, 
$\rho_{\rm cl}/\rho_{\rm ic}$, is thus higher than $1/f_{\rm ic}$, and 
%\textit{contrast} between 
%inter-clump medium and clumps, \textbf{$\rho_{\rm ic}/\rho_{\rm cl}$, 
%is thus lower} than $f_{\rm ic}$, and 
can be obtained from Eqs.~\ref{Eq:rhom} and \ref{Eq:fcl}.

These generic expressions are used to calculate 
all opacities in our clumpy models. However, as 
now described, the calculation of the clump optical depth 
$\tau_{\rm cl}$ depends on the specific absorbing/emitting 
process. 

\subsection{Calculating the clump optical depth} 

The clump optical depth for \textit{continuum} absorption is 
(SPO14)
\begin{equation} 
  %\tau_{\rm cl}^{\rm cont}(h,f_{\rm cl},f_{\rm ic}) 
   \tau_{\rm cl} = \langle \chi_{\rm c} \rangle h \ (1-(1-f_{\rm vol})f_{\rm ic}),
    %\equiv \langle \chi_{\rm c} \rangle G_{\rm c},
  \label{Eq:taucl_cont} 
\end{equation} 
where the porosity-length $h \equiv l_{\rm cl}/f_{\rm vol}$ 
is the \textit{mean-free-path} between clumps of characteristic 
length scale $l_{\rm cl}$, and the volume filling factor is 
related to the clumping factor 
via eqns.~\ref{Eq:fcl} and \ref{Eq:fic}. We note {\it i)} that for a void inter-clump 
medium ($f_{\rm ic}=0$), eqn.~\ref{Eq:taucl_cont} depends 
exclusively on $h$ and recovers the previously obtained 
result $\tau_{\rm cl} = \langle \chi_{\rm c} \rangle h$ \citep{Owocki04, Sundqvist12},
and {\it ii)} that the above expression assumes statistically 
isotropic clump optical depths, as supported by the 
empirical porosity-studies of \citet{Leutenegger13, Herve13} and 
by the theoretical models of \citet{Dessart03, Sundqvist18}. 

For \textit{line} absorption, $\tau_{\rm cl}$ may be similarly
calculated from the mean line-integrated opacity
\begin{equation} 
  \langle \chi_l \rangle = \frac{ \pi e^2}{ m_e c} f_{\rm osc}
  \frac{n_{\rm f,l}}{f_{\rm cl}}  \left( 1- \frac{n_{\rm f,u}g_{\rm l}}{n_{\rm f,l} g_{\rm u}} \right) %\frac{1}{f_{\rm cl}} ,
    \label{Eq:taud_cl}
\end{equation} 
which yields for a frequency-normalized profile function $\phi_\nu$
%and characteristic clump-length scale $l_{\rm cl}$ 
%
\begin{equation} 
  %\tau_{cl} =  \int_{l_{\rm cl}} \ \langle \chi_l \rangle
  %\frac{(1-(1-f_{\rm vol})f_{\rm ic})}{f_{\rm vol}}
  %\frac{\phi}{\Delta \nu_D} \ dl
  \tau_{cl} =  \int_{l_{\rm cl}} \ \langle \chi_l \rangle
  \frac{(1-(1-f_{\rm vol})f_{\rm ic})}{f_{\rm vol}}
  \phi_\nu \ dl
  \label{Eq:tau_cl}
\end{equation} 
where the integration extends over the clump and $\phi_\nu$ is 
evaluated at the corresponding co-moving frame (cmf) frequency; 
in hot stellar winds Doppler broadening normally dominates so 
that $\phi_\nu = \exp{(-x_{\rm cmf}^2)}
/\sqrt{\pi} \Delta \nu_D$ for Doppler width $\Delta \nu_D = (\nu_0/c) \varv_{th}$ 
and $x_{\rm cmf} = (\nu_{\rm cmf}-\nu_0)/\Delta \nu_D$. 
Other variables in eqns. \ref{Eq:taud_cl} and \ref{Eq:tau_cl} have 
their usual meanings. However, we remind that $f_{\rm osc}$ above denotes the 
oscillator strength (not to be confused here with the clumping factor 
$f_{\rm cl}$) and also that since $\langle \chi_l \rangle$ is 
integrated over the line-profile it has units of frequency 
over length (rather than just 1/length as for $\langle \chi_{\rm c} \rangle$).  

\paragraph{Sobolev approximation for line clump optical depth.} 
%\paragraph{In the supersonic atmosphere,} 
For the rapidly accelerating hot stellar winds in focus here, we may 
safely assume that clumps cover their resonance zones of a few
thermal widths $\varv_{\rm th} \approx 5-10 \, \rm km/s$ in velocity space 
\citep{Sundqvist10, Sundqvist11, Sundqvist14}.\footnote{This does not
include any phenomenological so-called "microturbulent" velocities in the wind, since such 
wind microturbulence typically is added in unclumped models 
in order to mimic just these effects of clumping; when synthesized directly 
from clumped LDI simulations, no ad-hoc microturbulence is 
required to model for example the extended black troughs of 
UV P-Cygni lines  (see Fig. 1 of \citealt{Sundqvist12c}).}   
In full analogy with the standard assumptions of the Sobolev 
approximation then \citep{Sobolev60}, we can
integrate eqn.~\ref{Eq:tau_cl} over the clump to obtain
%thus replace the clump length 
%scale in eqn.~\ref{Eq:tau_cl} with the radial ``clump Sobolev length'' 
%$l_{cl}^{Sob} \equiv v_{th}/(dv/dr)_{cl}$, whereby
%
\begin{equation} 
  \tau_{cl} = \langle \chi_l \rangle \frac{(1-(1-f_{vol})f_{ic})}{f_{vol}} 
    \frac{\lambda_0}{\varv'_{cl}} 
    = \frac{\tau_{\rm S}}{f_{\rm vor}}(1-(1-f_{vol})f_{ic}),  
    \label{Eq:tl} 
\end{equation} 
where $\tau_{\rm S} \equiv \langle \chi_{\rm l} \rangle \lambda_0/
\varv'_{\rm sm}$ is the radial Sobolev optical depth for the 
mean wind with spatial velocity gradient $\varv'_{\rm sm} = d\varv/dr_{\rm sm}$ 
and line rest wavelength $\lambda_0$. The last equality 
of eqn. \ref{Eq:tl} further uses the definition by SPO14 for the 
velocity clumping factor (see their eqn. 9 and Fig. 2)  
\begin{equation}
  f_{\rm vor} \equiv \left| \frac{\delta \varv}{\Delta \varv} \right|
  =  f_{\rm vol} \ \delta \varv /\delta \varv_{\rm sm},   
  \label{Eq:fvel} 
\end{equation} 
for clump velocity span $\delta \varv$ and velocity separation 
between clump-centers $\Delta \varv$, and where $\delta \varv_{\rm sm}$ is 
the velocity-span the clump \textit{would have} if it followed the 
underlying smooth velocity field. 
Note here that the special case of a clumped wind in which all 
clumps indeed follow this smooth velocity field, i.e. 
$\delta \varv = \delta \varv_{\rm sm}$, implies $f_{\rm vor} = f_{\rm vol}$ and 
so in general high clump optical depths ($\tau_{\rm cl} \approx \tau_{\rm Sob}/f_{\rm vol}$) 
and thus very large velocity-porosity effects; under typical circumstances 
though, $f_{\rm vor} > f_{\rm vol}$ (\citealt{Sundqvist10}, see also \S4). 

The un-normalized velocity clumping factor is related to the normalized, and perhaps 
somewhat physically more intuitive, velocity \textit{filling} factor, 
$f_{\rm vel}$, via 
\begin{equation} 
  f_{\rm vel} \equiv \frac{\delta \varv}{\delta \varv + \Delta \varv} = \frac{f_{\rm vor}}{1 +f_{\rm vor}}.  
\label{Eq:fvel_1} 
\end{equation} 
Eqns. \ref{Eq:rhom}-\ref{Eq:fvel_1} provide a full formalism toward implementing the effects of porosity in physical and velocity space into global NLTE model atmospheres like {\sc fastwind}, using the four parameters $f_{\rm cl}$, $f_{\rm ic}$, $h$, and $f_{\rm vel}$.

In {\sc fastwind}, we use this particular set of independent parameters because we believe they best represent the underlying physics of the situation. In principle, however, using the equations above one can simply substitute $f_{\rm cl}$, $h$, and $f_{\rm vel}$, for, e.g., $f_{\rm vol}$, $\ell_{\rm cl}$, and the clump velocity span $\delta \varv$. The key point though, is that four independent parameters are needed to fully describe the radiative transfer effects in the assumed two-component structured, accelerating medium. 

Nonetheless, as further discussed in \S4-5, depending on which diagnostic feature in which wavelength-regime is targeted, not all four parameters are always important in practice. We also note here that $f_{\rm cl}$, $f_{\rm ic}$ and $h$ are
equivalents to the parameters used to describe continuum transport in a static two-component, stochastic Markovian mixture model (see book by \citealt{Pomraning91}, and also SPO14), whereas the last, $f_{\rm vel}$, accounts here for the additional effects of spectral line transport in accelerating media (SPO14). 
%before in the next subsection describing our practical implementation of the formalism into {\sc fastwind}.  

\subsection{Practical implementation into {\sc fastwind}}

This section summarizes some key points regarding our 
practical implementation of the above formalism into {\sc fastwind}. 
The following radially dependent (but frequency independent) 
 auxiliary quantities are pre-calculated within {\sc fastwind} by a 
subroutine {\sc clumping}: 
\begin{equation} 
	G_{\rm c} \equiv  \ \left( 1-(1-f_{\rm vol})f_{\rm ic} \right) h, 
\end{equation} 
\begin{equation} 
G_{\rm l} \equiv  \left( 1-(1-f_{\rm vol})f_{\rm ic} \right) \frac{1-f_{\rm vel}} {f_{\rm vel} \varv_{\rm sm}'},   
\end{equation} 
where the clump volume filling factor is obtained from (the input parameters) 
$f_{\rm cl}$ and $f_{\rm ic}$ according to 
\begin{equation} 
	f_{\rm vol} = \frac{(1-f_{\rm ic})^2}{f_{\rm cl} - 2 f_{\rm ic} + f_{\rm ic}^2}. 
\end{equation}
Then, the \textit{frequency-dependent} clump optical depths and effective 
opacities are computed within the code's NLTE and radiative transfer networks
from:
\begin{equation} 
	\tau_{\rm cl,c} = \langle \chi_{\rm c} \rangle \ G_{\rm c}, 
\end{equation} 
\begin{equation} 
%	\tau_{\rm cl} = \tau_{\rm Sob} G_{\rm l}, 
	\tau_{\rm cl,l} =  \langle \chi_{\rm l} \rangle \lambda \ G_{\rm l}, 
\end{equation} 
\begin{equation} 
  \chi_{\rm eff} = \langle \chi \rangle \frac{1 + \tau_{\rm cl} f_{\rm ic}}{1+\tau_{\rm cl}}.  
  %\equiv  \langle \chi \rangle \, f_{\rm red}(\tau_{\rm cl}, f_{\rm ic}),
\label{Eq:chi_e} 
\end{equation}    
where subscripts 'c' and 'l', denoting continuum and line, have been omitted for simplicity 
in eqn.~\ref{Eq:chi_e}, and where all mean opacities
include corrections for clumping 
(cf. eqns.\ref{Eq:taucl_cont} and \ref{Eq:taud_cl}), with occupation numbers 
derived for density $\rho_{\rm f} = \rho_{\rm sm} f_{\rm cl}$. 

\paragraph{Input.} Four structure-parameters, the clumping factor $f_{\rm cl}$, 
the porosity length $h$, the inter-clump density contrast parameter 
$f_{\rm ic}$, and the velocity filling factor $f_{\rm vel}$, are 
provided by the user as input. As described further below, 
these parameters are typically functions of radius
(also controlled by user). If one chooses to pre-specify only $f_{\rm cl}$, 
we force all clump optical depths to be effectively zero, and so 
recover previous models assuming optically thin clumps\footnote{We 
note that by rescaling the relation between $f_{\rm cl}$ and 
$f_{\rm vol}$, such models assuming optically thin 
clumps do not necessarily imply an effectively void inter-clump medium 
(see previous section).}. Similarly, simply setting $f_{\rm cl}=1$ means the two components 
are identical so that a smooth wind model with density 
$\rho_{\rm sm}$ is calculated, and the same is true also 
if $f_{\rm ic} =1$ is provided.  

\paragraph{Velocity-space porosity in supersonic part of wind.} As outlined 
above, we essentially use the Sobolev approximation to evaluate the clump 
optical depths for spectral lines. This method is only motivated for the 
supersonic parts of the wind\footnote{Or more accurately for the 
"superthermal" parts $\varv > \varv_{\rm th}$ for the considered ion; to 
allow for a general treatment, however, we have chosen here simply  
$\rm max \it (\varv_{\rm th}) = a$.}. As such, the present version of {\sc fastwind} allows 
the consideration of such velocity-space porosity only for the 
supersonic parts $\varv > a$; to this end, we have for simplicity set the same lower 
boundary also for spatial porosity and optically thin 
clumping (i.e., we do not allow for any clumping in the sub-sonic parts).    

\paragraph{Radial variation of porosity length.} 
As default in {\sc fastwind}, we assume a radial variation 
$h = h_\infty \varv/\varv_\infty$, where $h_\infty$ is the porosity-length 
at terminal wind speed (provided by user). Such a 'velocity-stretch'
law is consistent with various other models
\citep[e.g.,][]{Oskinova04, Sundqvist12, Leutenegger13, 
Grinberg15}, and follows naturally from considering mass-conserving 
clumps with characteristic length-scales that expand 
according to the local wind radius \citep{Sundqvist12}. 
We note that while this law certainly is very reasonable as default, the user 
can readily modify it  (e.g., to account for collisional 
merging of clumps, \citealt{Feldmeier97}) from within the 
above-mentioned subroutine {\sc clumping}. 
    
\paragraph{Standard and background opacities.} The 
implementation of the above formalism into 
any `standard' method for 
solving the coupled radiative transfer and NLTE rate equations  
is straightforward; for each considered process, `smooth' or 
mean opacities are converted to `effective' ones using the above expressions for 
the (frequency dependent) mean opacities and the 
(frequency independent) $G$-quantities (defined above 
to obtain clump optical depths and effective opacities). Whenever 
more than one process is present at a given frequency, the 
total clump optical depth is computed by summing up all 
individual components, and the corresponding effective 
opacities computed using this summed $\tau_{\rm cl}$. After 
this, calculations proceed as usual, but now using the
new \textit{effective} opacities $\chi_{\rm eff}$ in the radiative transfer
(instead of as in previous models mean opacities $\langle \chi \rangle$).
We remind again here that the NLTE rate equations have to be solved 
for $\rho_{\rm f} = \rho_{\rm sm} f_{\rm cl}$.  

For the so-called background line opacities included in {\sc fastwind} (see 
previous section), the implementation becomes a bit more 
intricate. To this end, we proceed as follows: as explained in detail 
in \citet{Puls05}, these background line opacities are calculated by building  
suitable statistical averages within a number of pre-defined discrete 
frequency-bins. In each of these bins, we now also compute clump optical 
depths by summing up the opacities of the contributing line processes. These clump 
optical depths are then used to obtain an effective opacity for each 
of the bins; it is this effective opacity that then finally is used in all  
transfer calculations to obtain the line-blanketed pseudo-continuum 
radiation field. Whenever there is an overlap between these 
background lines and a line treated "explicitly"
(see above), we compute $\tau_{\rm cl}$ using both approaches 
and simply choose the highest one when calculating 
the effective opacity finally used in the radiative transfer. 
 
%\paragraph{Formal integral.} Finally, the implementation into the
%so-called formal integral of radiative transfer, used to compute synthetic 
%spectra, is trivial, and follows from scaling the mean opacities 
%of our NLTE model in the same manner as above, and then solving 
%the formal integral using the effective opacities. In practice, we
%compute a reduction factor $\chi_{\rm eff}/\langle \chi \rangle$ for each 
%radius and wavelength (for both background and line-specific 
%opacities), and then simply scale the final opacities 
%entering the radiative transfer equation.  

\section{Tests and first results} 

\begin{figure*}
  %  \vspace{1cm}
    	\begin{minipage}{6.3cm}
      \resizebox{\hsize}{!}  {\includegraphics[]{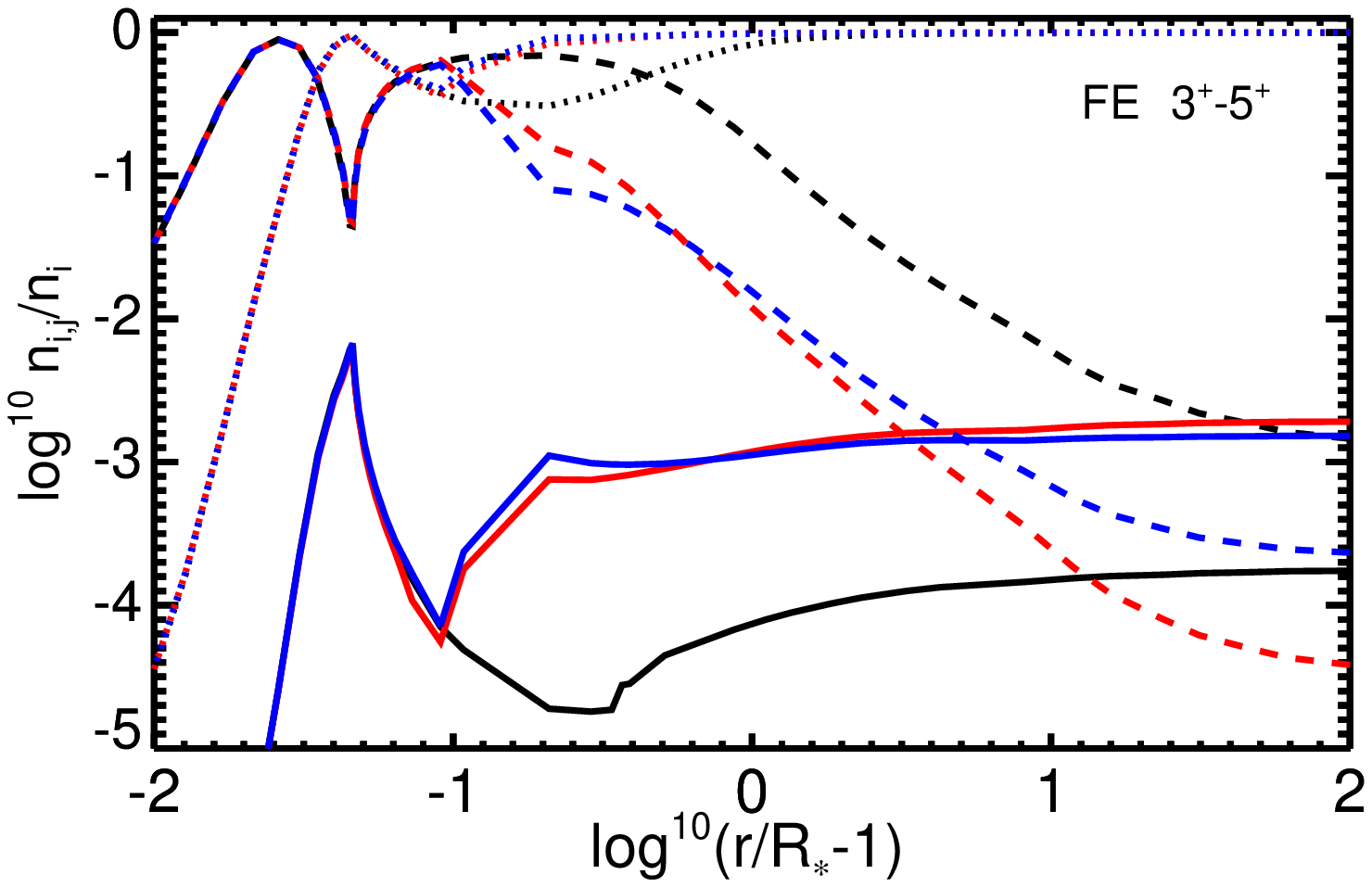}}
    \centering
   \end{minipage}
       	\begin{minipage}{6.3cm}
      \resizebox{\hsize}{!}  {\includegraphics[]{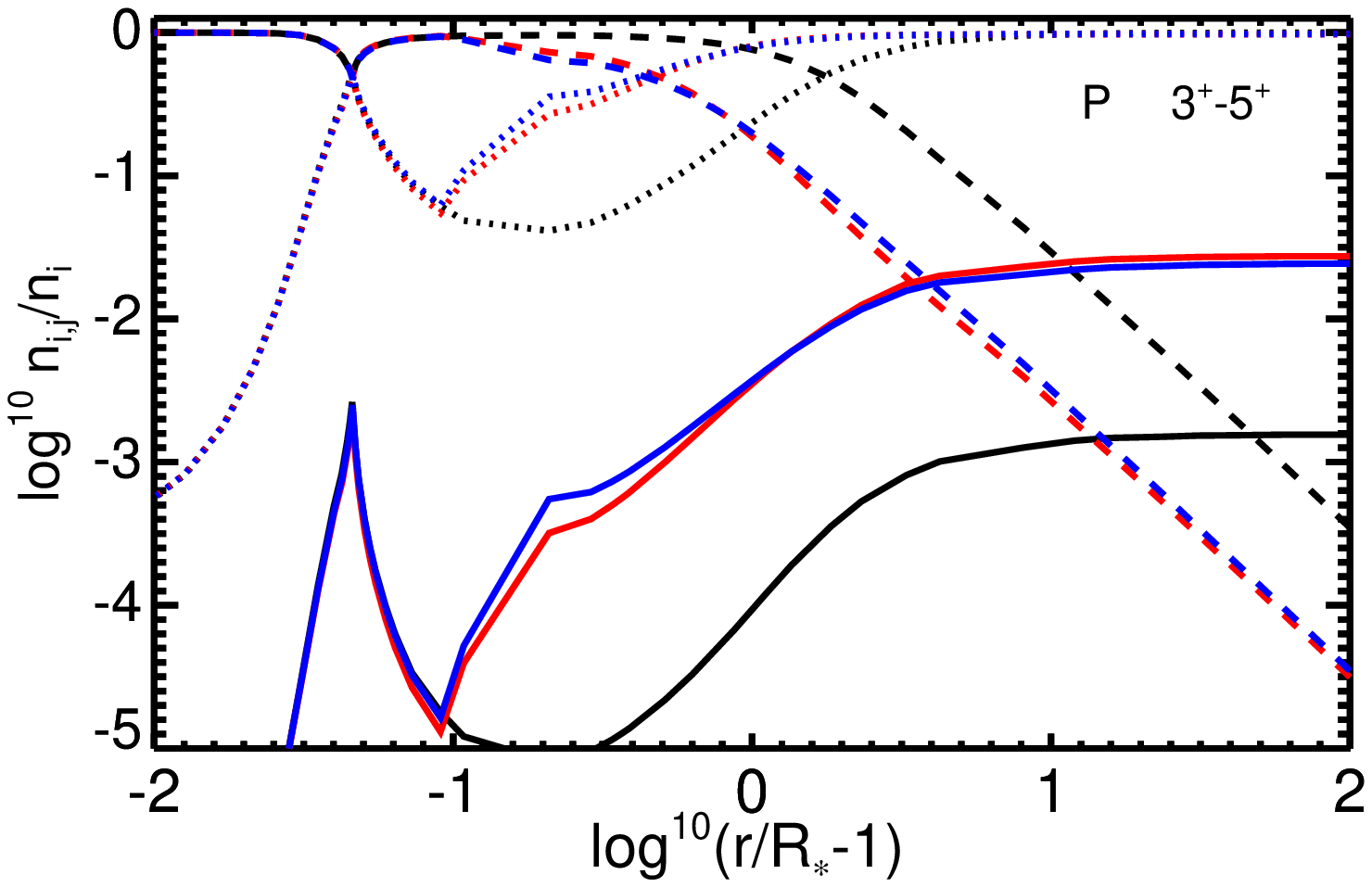}}
    \centering
   \end{minipage}
       	\begin{minipage}{6.3cm}
      \resizebox{\hsize}{!}  {\includegraphics[]{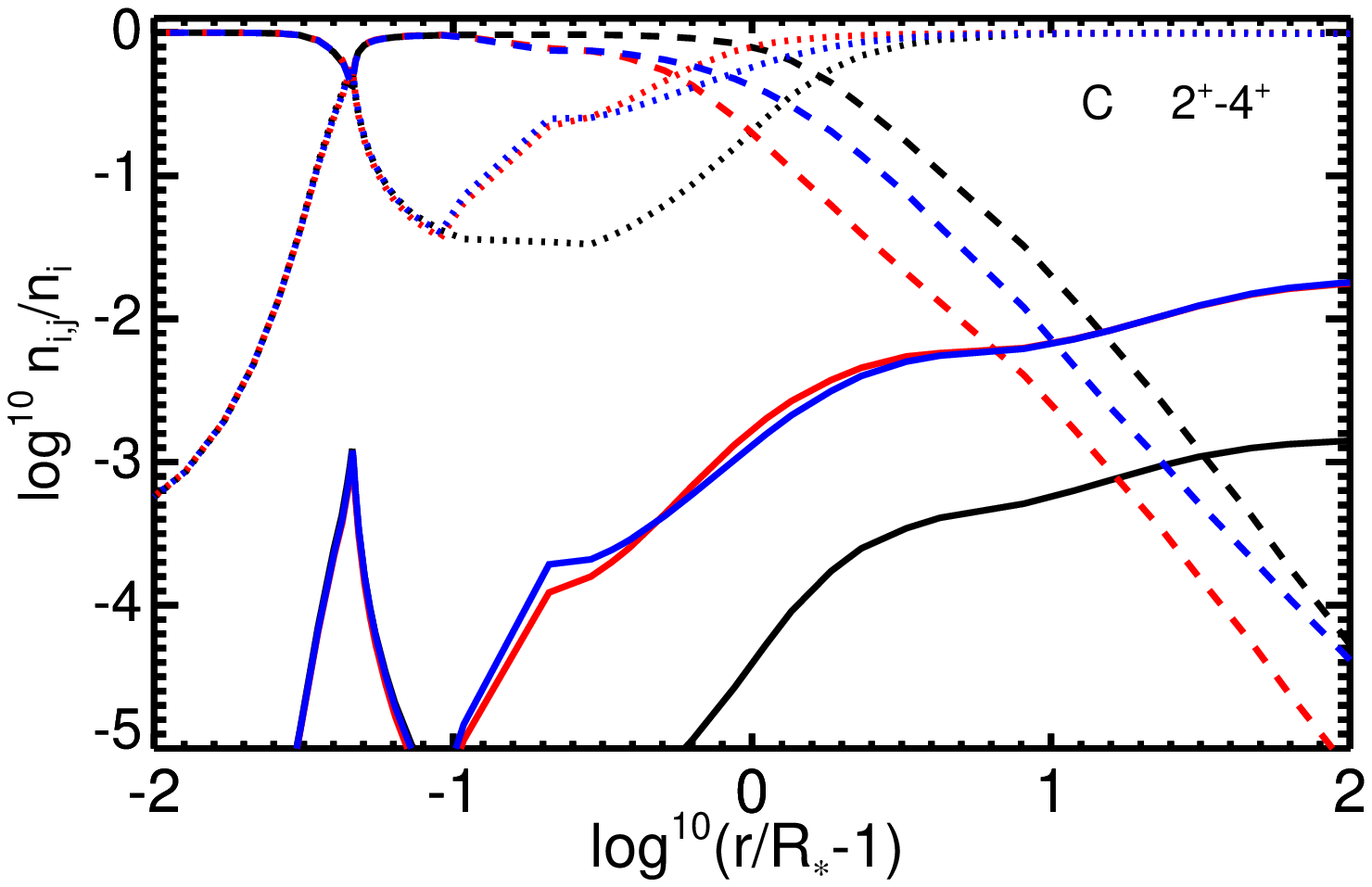}}
    \centering
   \end{minipage}
       	\begin{minipage}{6.3cm}
      \resizebox{\hsize}{!}  {\includegraphics[]{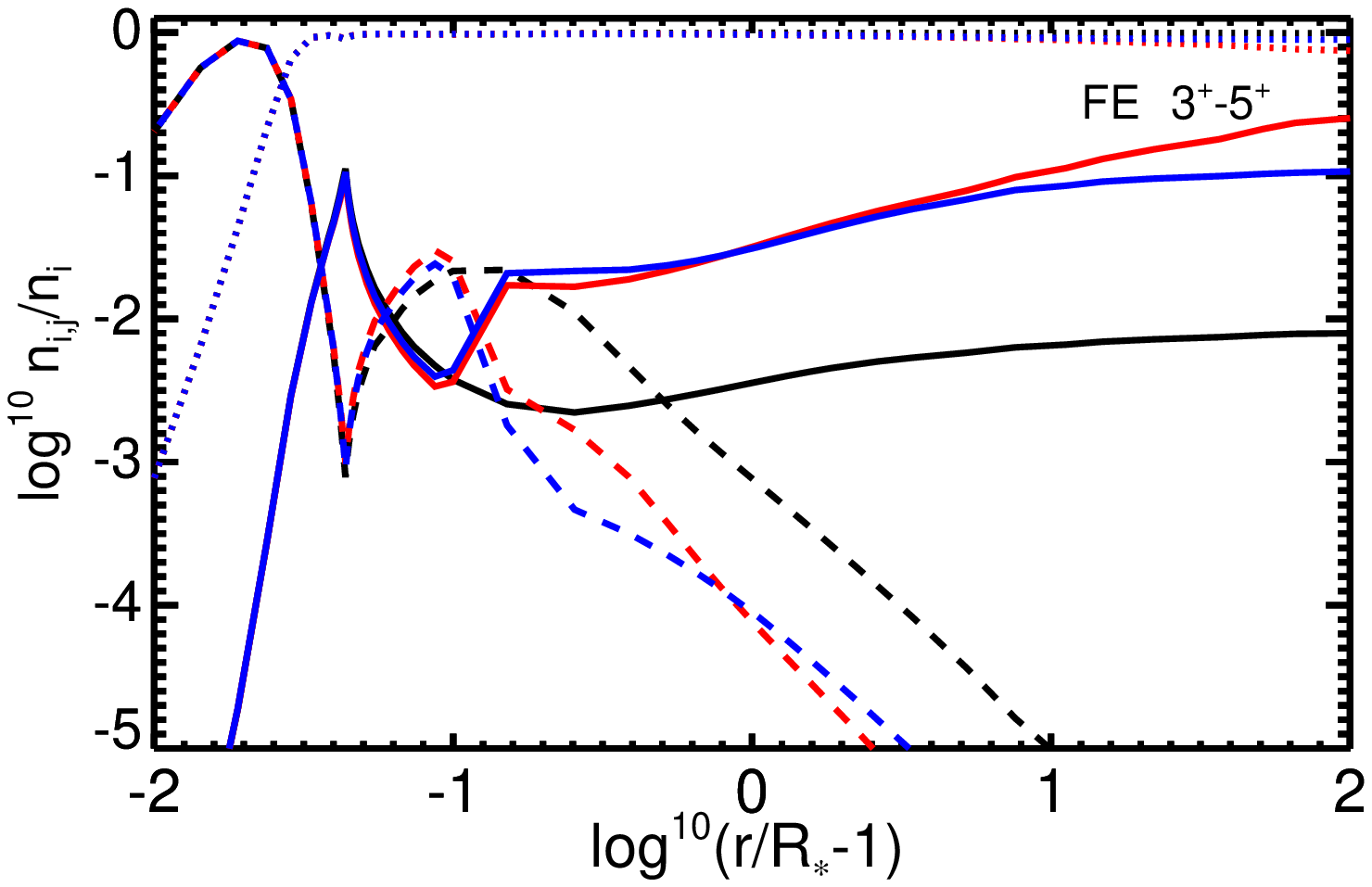}}
    \centering
   \end{minipage}
       	\begin{minipage}{6.3cm}
      \resizebox{\hsize}{!}  {\includegraphics[]{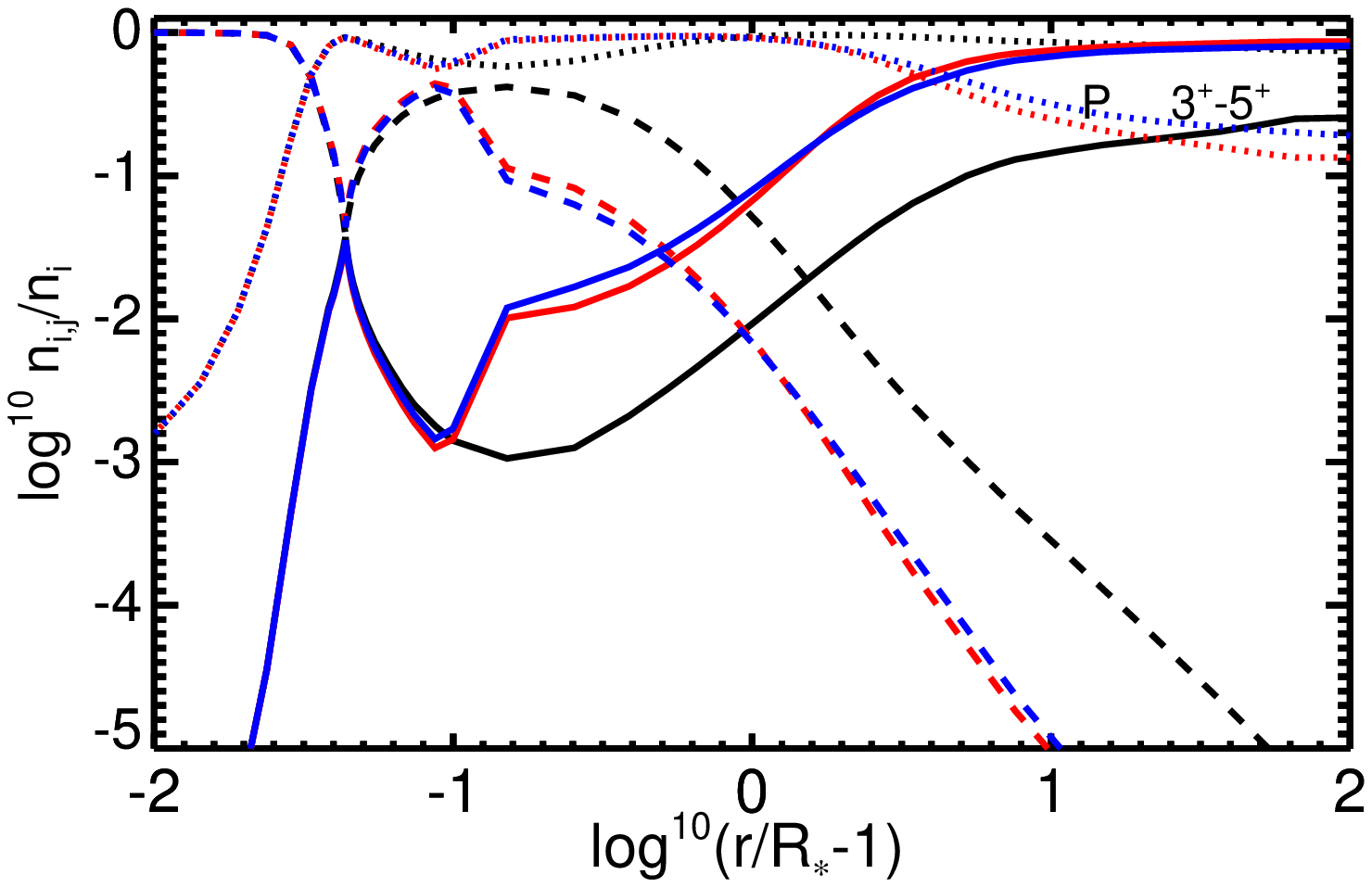}}
    \centering
   \end{minipage}
       	\begin{minipage}{6.3cm}
      \resizebox{\hsize}{!}  {\includegraphics[]{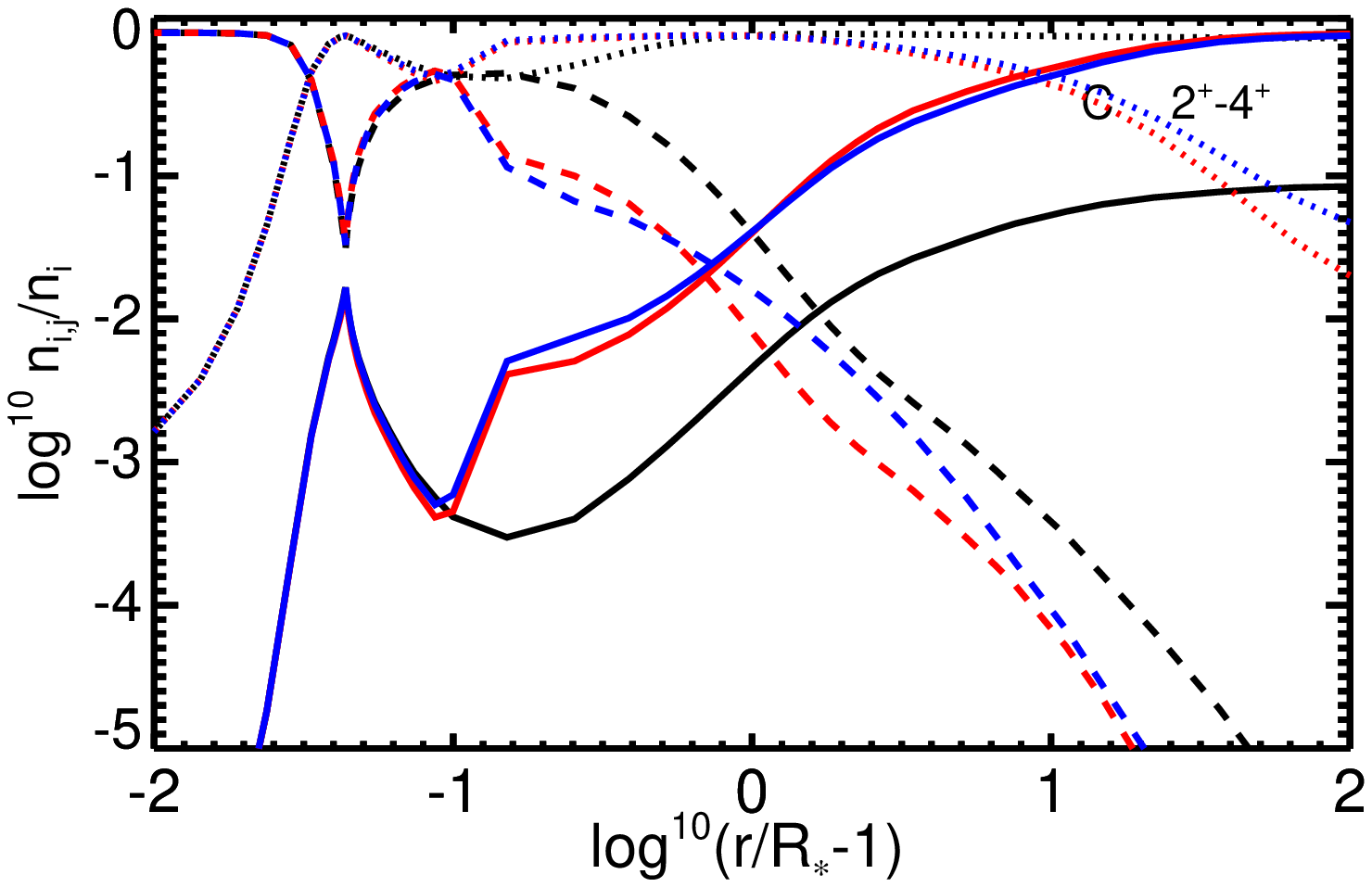}}
    \centering
   \end{minipage}
          	\begin{minipage}{6.3cm}
      \resizebox{\hsize}{!}  {\includegraphics[]{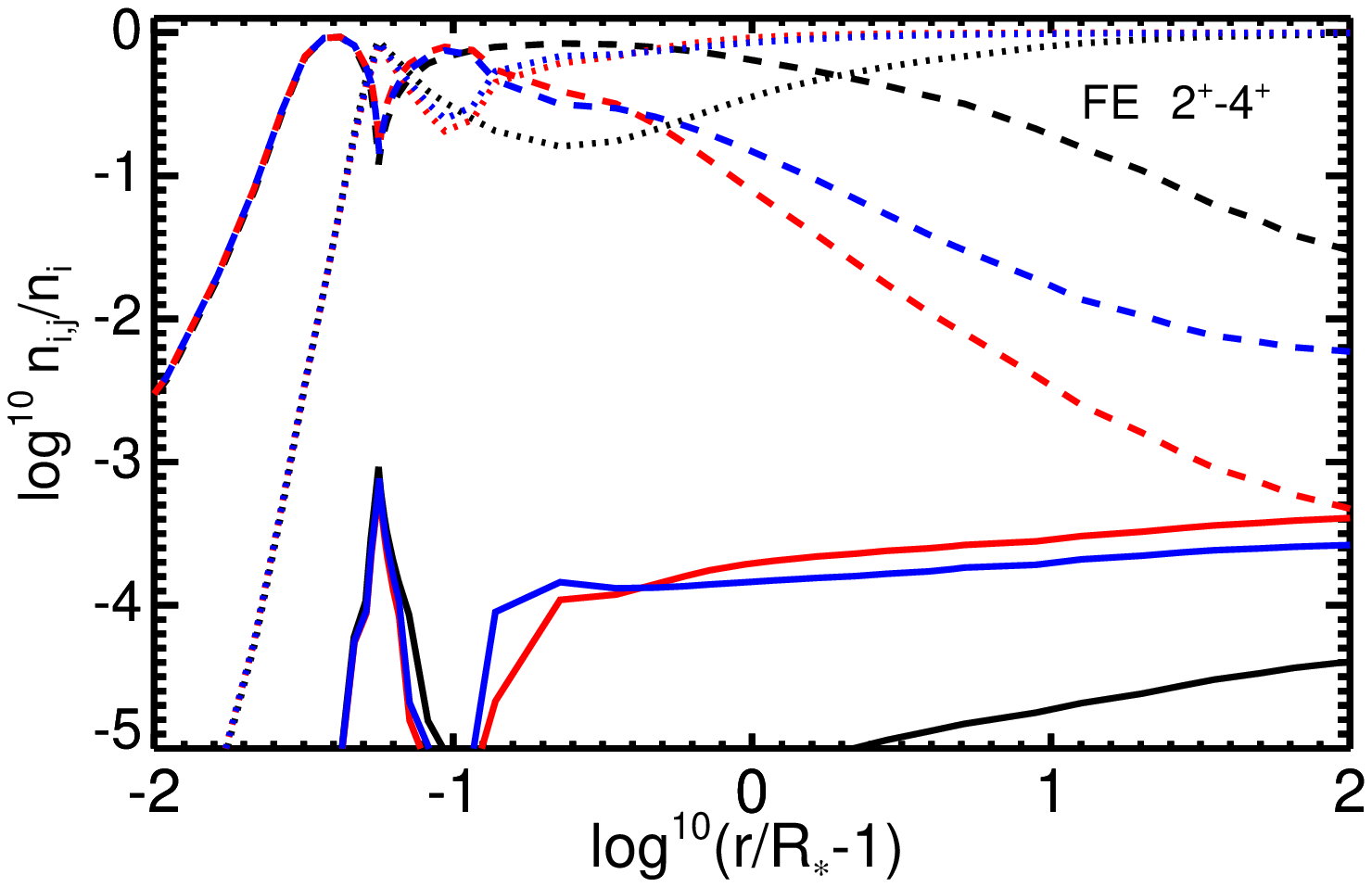}}
    \centering
   \end{minipage}
       	\begin{minipage}{6.3cm}
      \resizebox{\hsize}{!}  {\includegraphics[]{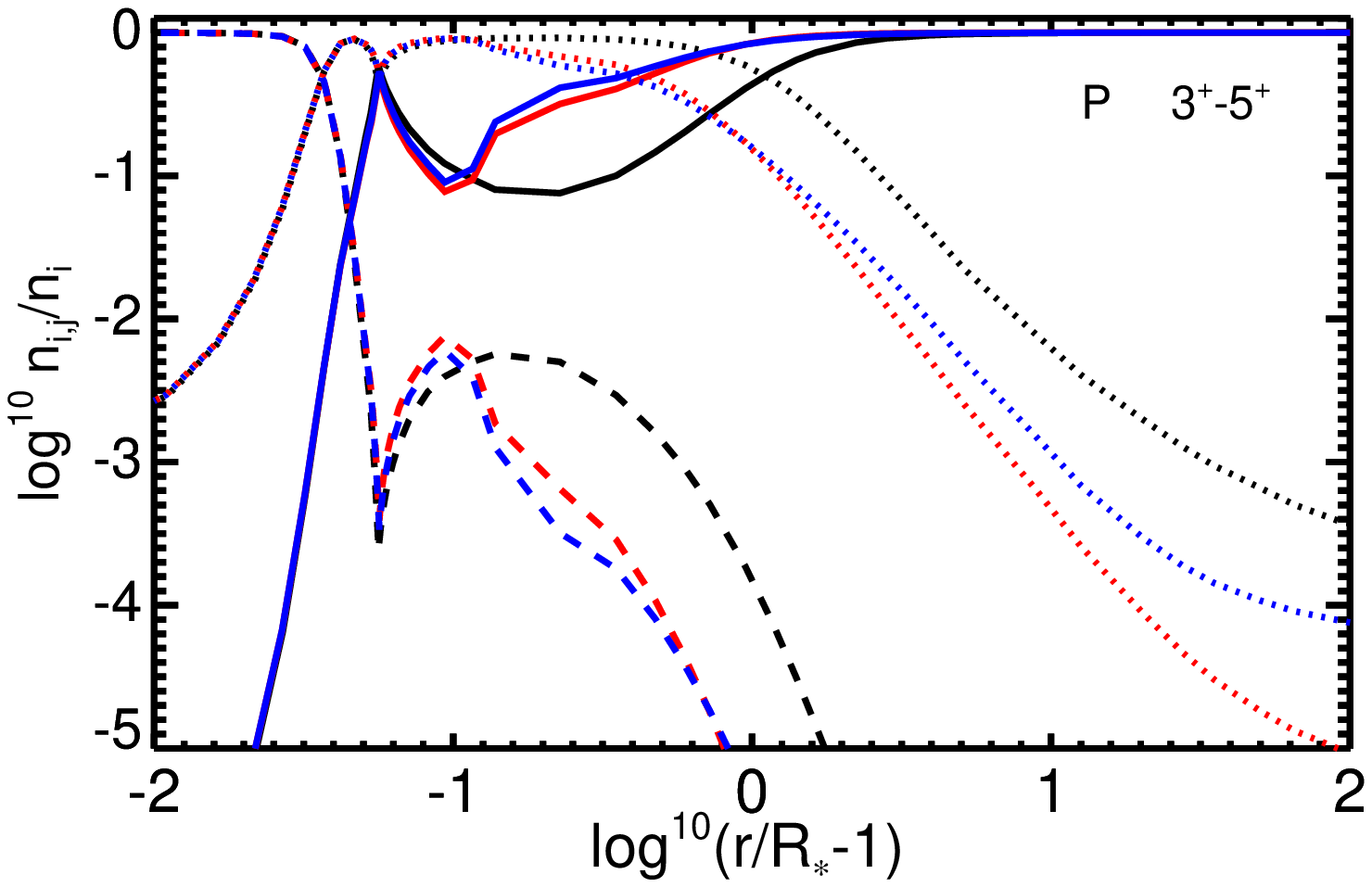}}
    \centering
   \end{minipage}
       	\begin{minipage}{6.3cm}
      \resizebox{\hsize}{!}  {\includegraphics[]{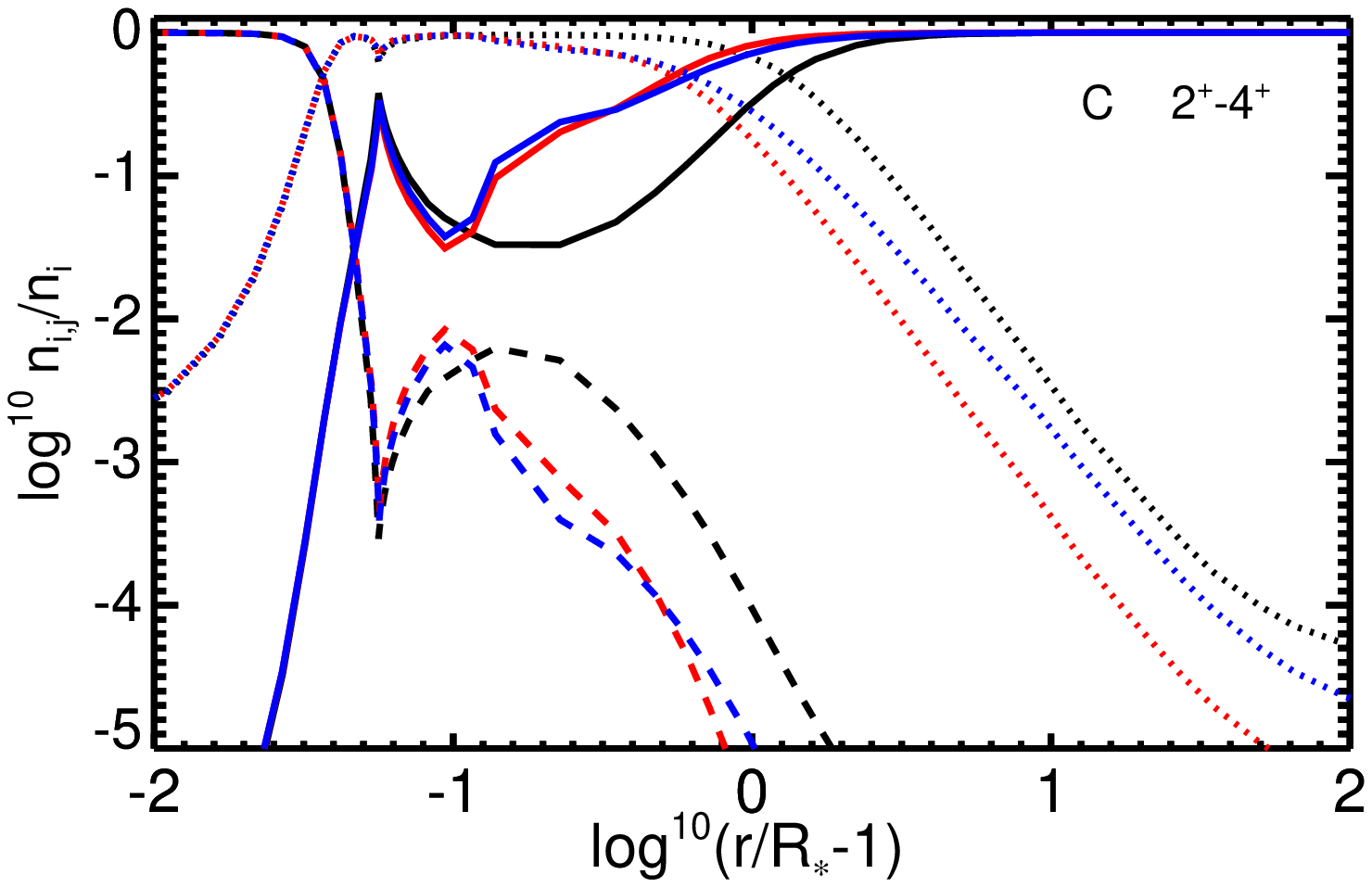}}
    \centering
   \end{minipage}
  \caption{Ionization balance of selected elements iron (Fe), phosphorus (P), and carbon 
  (C) for {\sc fastwind} models `O1' (top panels), `O2' (middle panels), and 
  `O3' (lower panels) with parameters according to Table 1. Plotted in each panel are  
  number density ratios $\log n_j/n_i$ for ion stage $j$ of species $i$, with 
  $n_i$ the total element abundance, 
  as function of a radius coordinate $\log^{10} (r/R_{*}-1)$.
  Blue lines are models 'thick' including full effects of porosity in physical and velocity space; red lines 
  are models 'thin' assuming optically thin clumping; black lines are models 'smooth' 
  without any wind clumping. For iron (left panels), phosphorus 
  (middle panels), and carbon (right panels), the displayed states are given in the upper right 
  of each panel; solid lines then show the ion fraction in the 
  lowest considered state $j$, the dotted lines show state 
  $j+1$ and the dashed ones $j+2$. (For example the top left panel thus shows Fe\,$3^{+}-5^{+}$ = Fe\,$\rm IV-VI$ in models O1 'thick' (blue), 
  'thin' (red) and 'smooth' (black), where solid lines are 
  ion fractions of Fe IV, dotted lines of Fe V, and dashed lines of Fe VI.)}  
  \label{Fig:ion}
\end{figure*}

\begin{figure}
  %  \vspace{1cm}
  %  	\begin{minipage}{16.0cm}
      \resizebox{\hsize}{!}  {\includegraphics[]{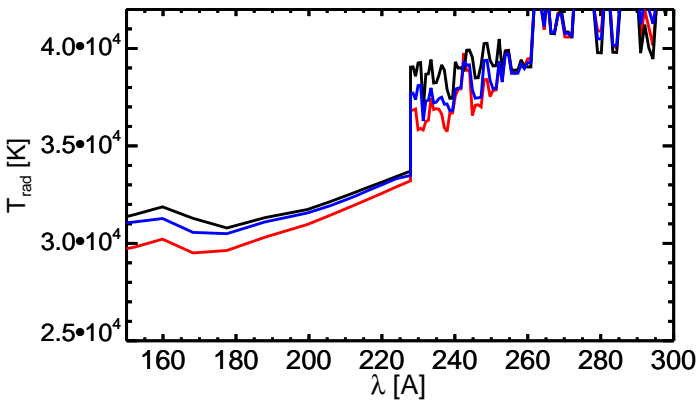}}
    \centering
 %   \end{minipage}
  \caption{Radiation temperature $T_{\rm rad}$ vs. wavelength at a wind radius $r/R_\ast \approx 2$ for 
  the three 'O1' models in Table 1. Models are colored as in Fig. 1, i.e. the blue line is model 'thick', the red 
  line model 'thin' and the black line model 'smooth'.}  
  \label{Fig:trad}
\end{figure}

\begin{figure}
  %  \vspace{1cm}
  %  	\begin{minipage}{16.0cm}
      \resizebox{\hsize}{!}  {\includegraphics[]{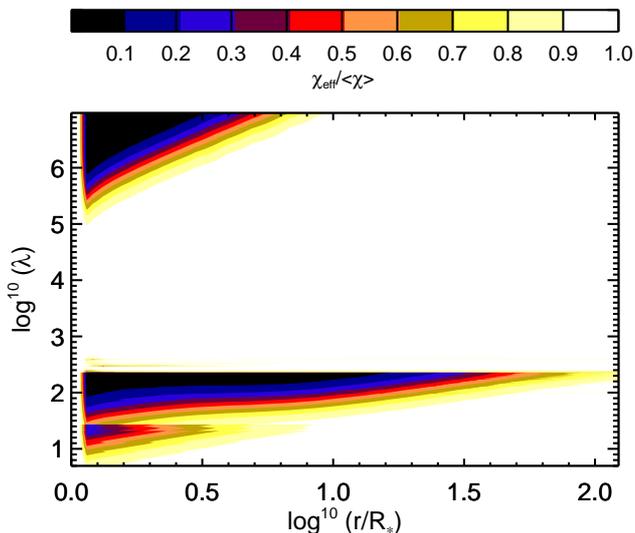}}
    \centering
 %   \end{minipage}
  \caption{Contour plot of effective to mean opacity ratio, $\chi_{\rm eff}/\langle \chi \rangle$, for complete 
  background opacities (i.e., sum of background lines + total continuum, see text), 
  displayed as function of radius (abscissa) $r/R_\ast = 1-100$ and the full considered wavelength range (ordinate) 
  between 5 and $10^7$ $\AA$. Colors range between no opacity-reduction $1$ (white) and the limiting value $f_{\rm ic} = 0.01$ 
  (black). Parameters according to $\zeta$ Pup like 
  model 'thick1' in Table 1.}  
  \label{Fig:kappa_eff}
\end{figure}
 
\begin{figure}
%    \vspace{1cm}
    	\begin{minipage}{ 8.0cm}
      \resizebox{\hsize}{!}  {\includegraphics[]{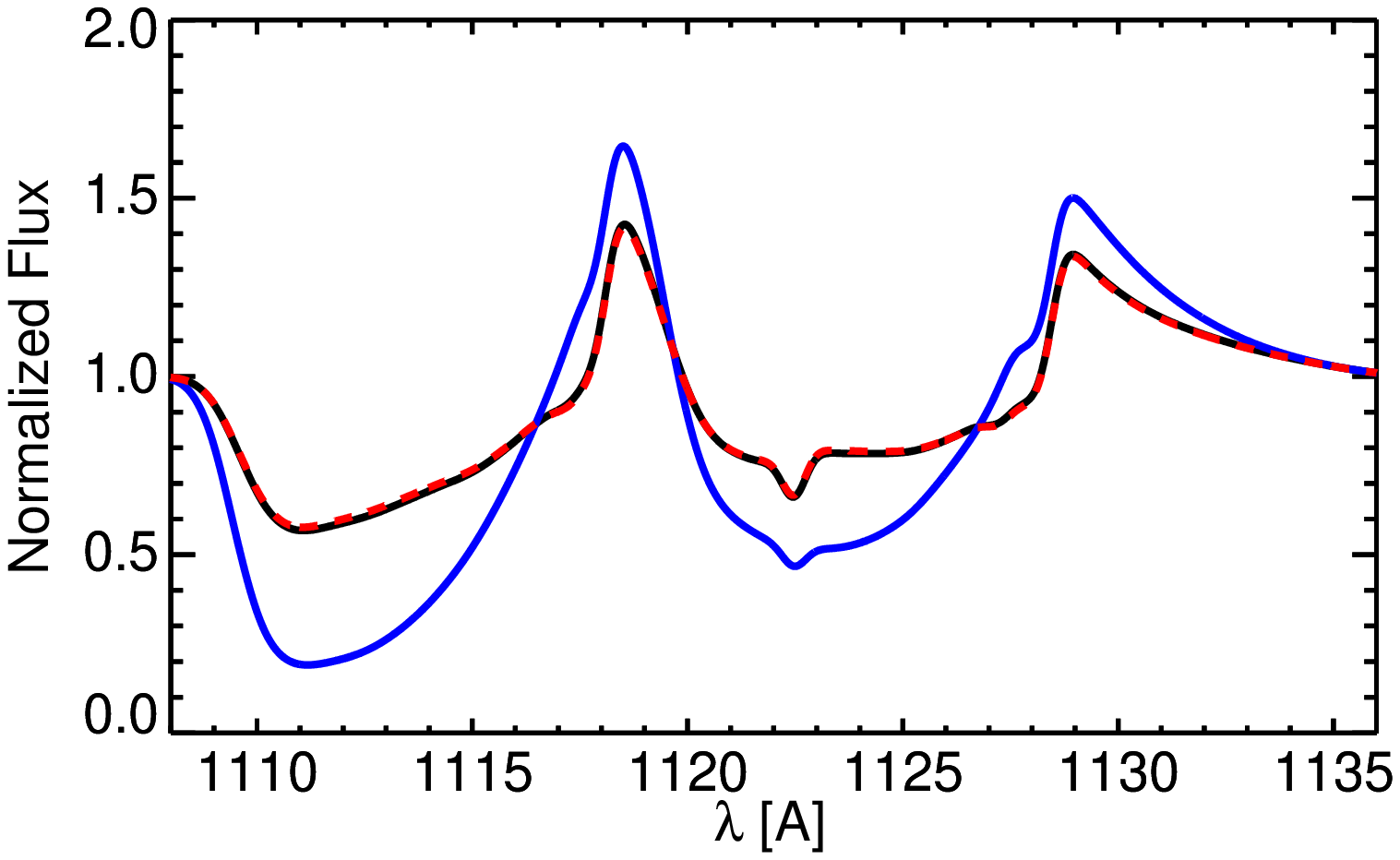}}
    \centering
    \end{minipage}
    
       \begin{minipage}{ 8.0cm}
      \resizebox{\hsize}{!}  {\includegraphics[]{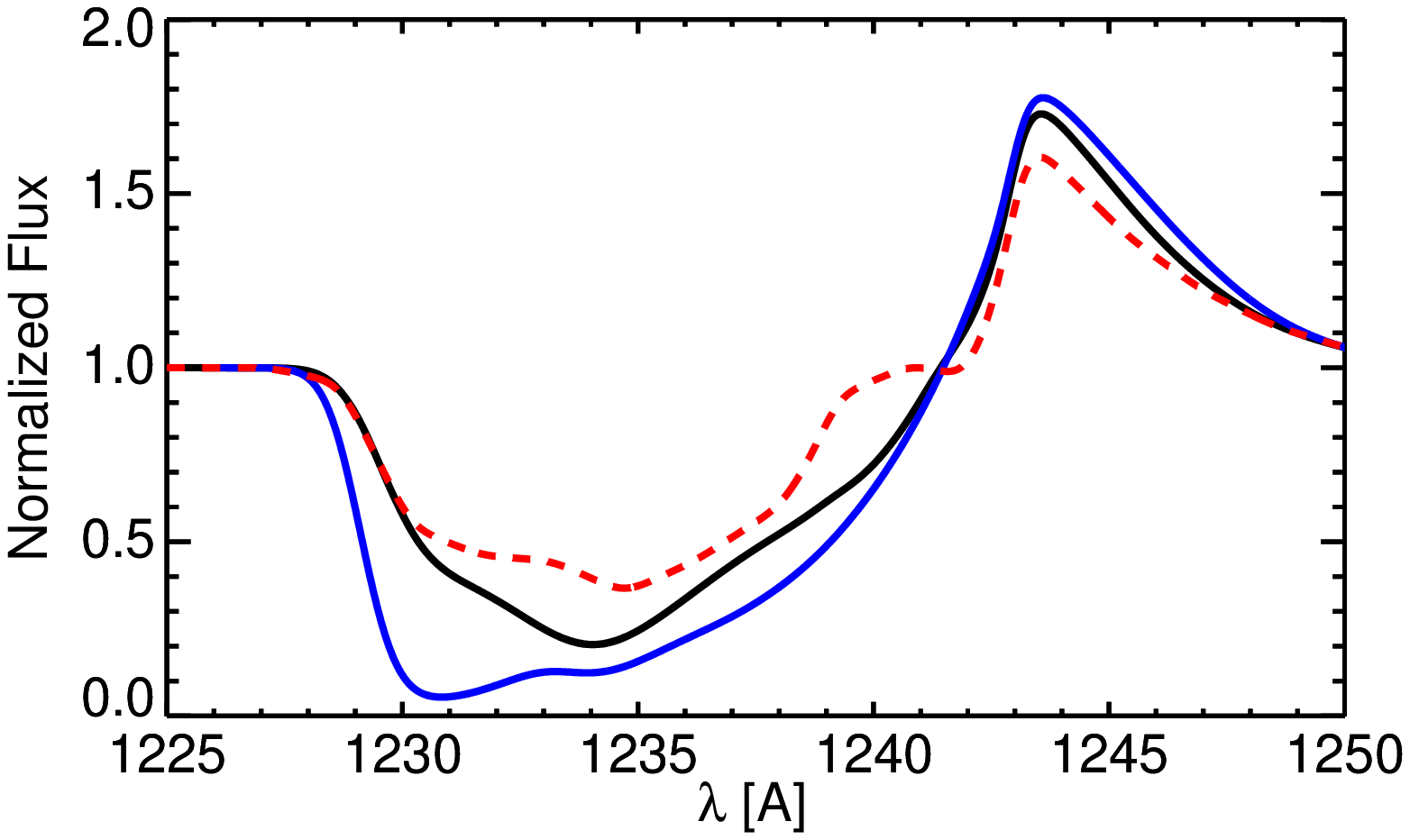}}
    \centering
    \end{minipage}
  \caption{Normalized flux profiles vs. wavelength for UV 'P-Cygni' resonance 
  doublets of PV (upper panel) and NV (lower panel). Profiles from 
  the three $\zeta$ Pup like models in Table 1 are shown, where blue lines display model 'thin', black   
  lines model 'thick1', and red dashed lines model 'thick2'. See text.}  
  \label{Fig:uv}
\end{figure}

\begin{figure}
%    \vspace{1cm}
    	\begin{minipage}{7.0cm}
      \resizebox{\hsize}{!}  {\includegraphics[]{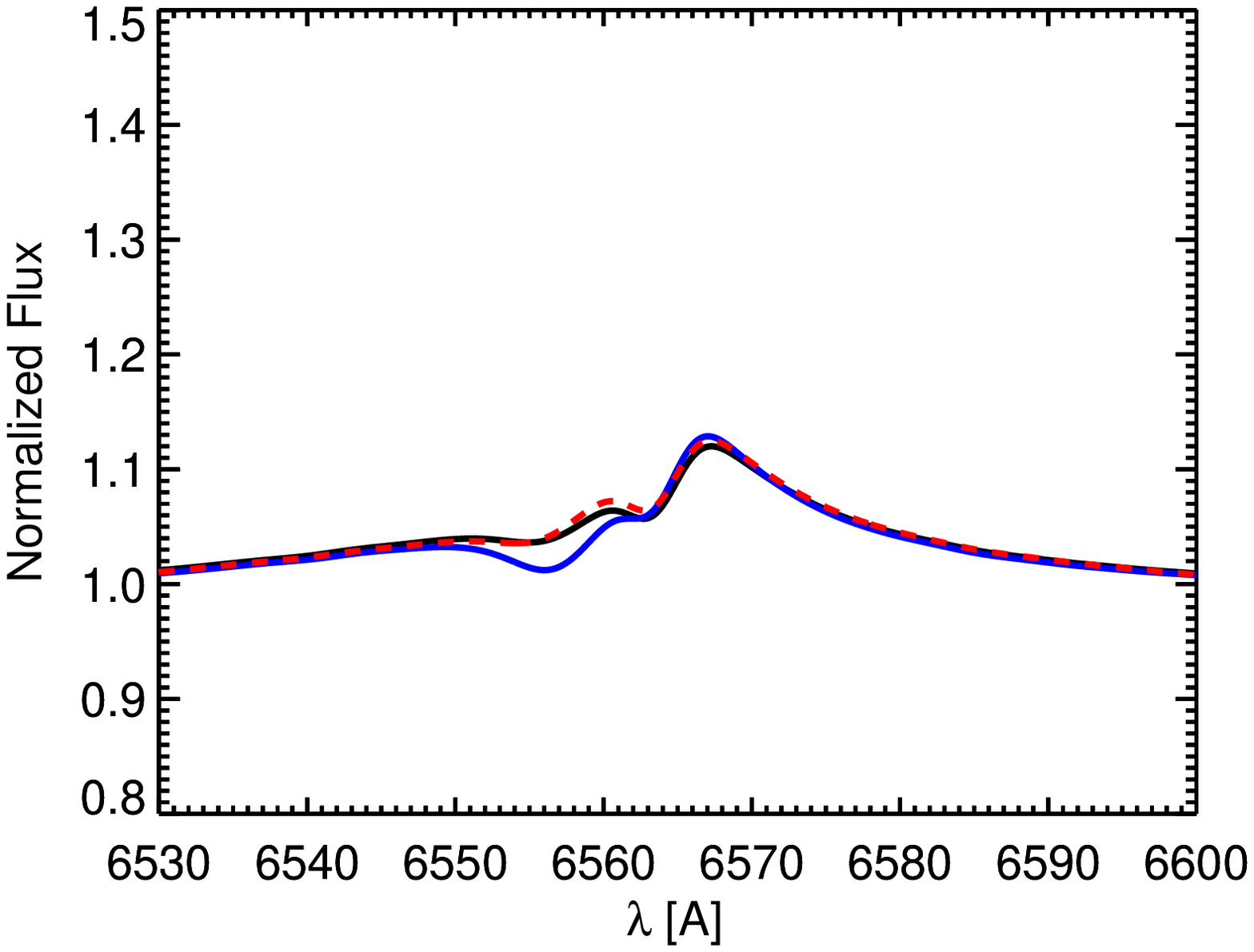}}
    \centering
    \end{minipage}
    
       \begin{minipage}{7.0cm}
      \resizebox{\hsize}{!}  {\includegraphics[]{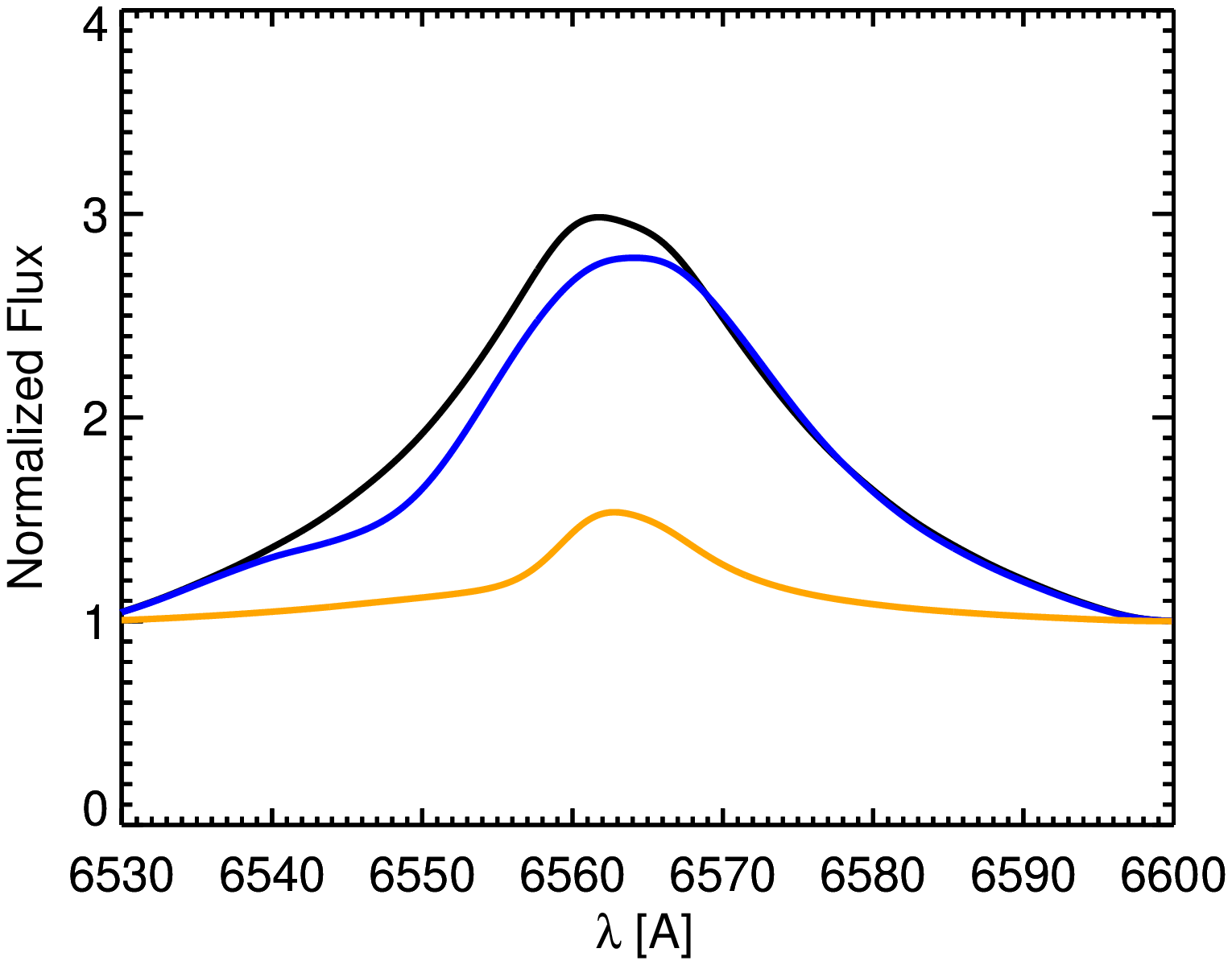}}
    \centering
    \end{minipage}
    
       \begin{minipage}{7.0cm}
      \resizebox{\hsize}{!}  {\includegraphics[]{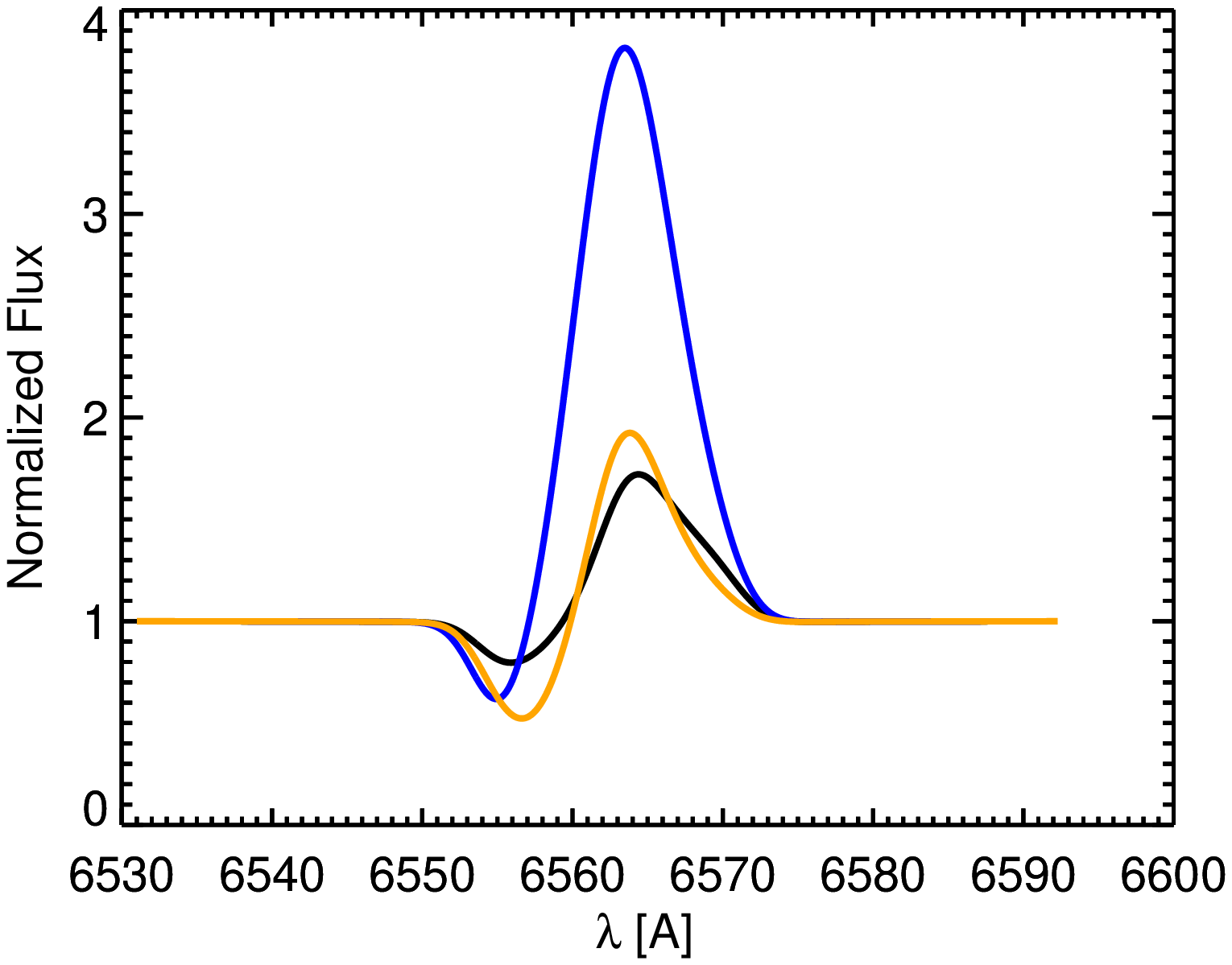}}
    \centering
    \end{minipage}
    
  \caption{Modeled H$\alpha$ line profiles. Profiles from 
  the three $\zeta$ Pup like models in Table 1 are shown in the upper panel, 
  with color coding as in Fig.~\ref{Fig:uv}. The middle and lower panels then 
  display the 'O3' and 'B7'
  models in Table 1, respectively, with black lines showing models 'thick', blue 
  lines models 'thin' and orange lines models 'smooth'.}  
  \label{Fig:ha}
\end{figure}

Let us now build physical insight by examining some important limits and basic results 
of the clumping model developed in \S3.

\subsection{Analytic limits} 

\paragraph{Smooth conditions.} In the formalism above, 
setting $f_{\rm ic} = 1$ (i.e., assuming two identical components) 
gives $\chi_{\rm eff} = \langle \chi\rangle$ and $f_{cl} =1$, thus
recovering the smooth background model. 
 
\paragraph{Optically thin clumps.} 
For optically thin clumps, $\tau_{\rm cl} \ll 1$, eqn.~\ref{Eq:chi_eff} 
always gives $\chi_{\rm eff} = \langle \chi \rangle$. 

For a void 
inter-clump medium, $f_{\rm ic} = 0$, this then results in
\begin{equation} 
 % f_{\rm cl} = \frac{1 - f_{\rm ic} (1-f_{\rm vol})(2- f_{\rm ic} )}{f_{\rm vol}} 
 % = 1/f_{\rm vol} 
   f_{\rm cl} = 1/f_{\rm vol} 
\end{equation} 
and 
\begin{equation} 
  \langle \chi \rangle \propto \langle \rho \rangle f_{\rm vol} / f_{\rm vol} 
  = \langle \rho \rangle,    
\end{equation} 
\begin{equation} 
  \langle \chi \rangle \propto \langle \rho \rangle^2 f_{\rm vol} / f_{\rm vol}^2 
  = \langle \rho \rangle^2 / f_{\rm vol} =  \langle \rho \rangle^2 f_{\rm cl},  
\end{equation} 
for processes depending on $\rho$ and $\rho^2$, respectively. This is consistent 
with previous {\sc fastwind} versions (as well as with alternative 
NLTE global atmospheres codes like {\sc cmfgen}), and 
further demonstrates that in this case the fiducial clump density  
$\rho_f = \langle \rho \rangle f_{cl} = \langle \rho \rangle/f_{vol} = \rho_{cl}$ 
indeed is the ``real'' clump density; that is, in models assuming 
an effectively void inter-clump medium the rate equations are solved 
within the clumps exactly. 

Adding a non-negligible inter-clump medium to this changes the relation between the 
clumping and volume filling factor, 
\begin{equation} 
  f_{\rm cl} = \frac{1 - f_{\rm ic} (1-f_{\rm vol})(2- f_{\rm ic} )}{f_{\rm vol}},  
\end{equation} 
but, \textit{when expressed in terms of the clumping factor}, preserves 
the scalings for $\rho$ and $\rho^2$ dependent opacities, 
\begin{equation} 
  \langle \chi \rangle \propto \langle \rho \rangle f_{\rm cl} / f_{\rm cl} 
  = \langle \rho \rangle,    
\end{equation} 
\begin{equation} 
  \langle \chi \rangle \propto \langle \rho \rangle^2 f_{\rm cl}^2 / f_{\rm cl} 
  =  \langle \rho \rangle^2 f_{\rm cl}.  
\end{equation} 
This illustrates the advantage of using the clumping factor instead of the volume 
filling factor as an independent parameter. Note though, that in this case of 
$f_{\rm ic} > 0$ {\it i)} the clumping factor does not exactly equal the overdensity 
of clumps (i.e., $\rho_{\rm cl} \ne \langle \rho \rangle f_{\rm cl}$) and {\it ii)} 
the inverse of the clumping factor does not precisely equal the volume filling factor. 
%, but rather
%
%\begin{equation}
%  $\rho_f = \langle \rho \rangle f_{cl}
%\end{equation} 

\paragraph{Optically thick clumps.} 
Let us first consider the case of very optically thick clumps, $\tau_{cl} \gg 1$, and 
a void inter-clump medium, $f_{\rm ic} = 0$. 

For spectral lines, this limit is best illustrated by an effective optical depth in the Sobolev approximation,  
\begin{equation} 
	 \tau_{\rm eff} \approx \langle \chi_{\rm eff} \rangle \lambda / \varv_{\rm sm}' = f_{\rm vor}
	 =\frac{f_{\rm vel}}{1-f_{\rm vel}},  
	 \label{Eq:fvel} 
\end{equation} 
which shows explicitly that the absorption here is controlled by how much of the 
wind velocity law is covered by clumps, i.e. by the velocity filling factor. 

For continuum processes, on the other hand, it is more instructive to consider
directly the effective opacity, for which one obtains \citep[e.g.,][]{Feldmeier03, Owocki04} 
\begin{equation} 
	 \chi_{\rm eff} \approx 1/h.  	
\end{equation} 
This equation now illustrates how in this case the opacity (and thus the optical depth) 
is set by the mean free-path between clumps, that is by the porosity length. 

Note that in these cases with $f_{\rm ic}=0$ the optical depth of the medium 
becomes \textit{gray} (frequency independent), both for lines and continua, 
and further how the medium's effective absorption 
becomes \textit{independent} of the atomic opacity 
(i.e., independent of occupation numbers and cross sections).  

Adding to this now a tenuous, but non-void, inter-clump medium (still considering 
optically thick clumps) yields for lines (SPO14) 
\begin{equation} 
 \tau_{\rm eff} \approx f_{\rm vor} + \tau_{\rm S} f_{\rm ic} = f_{\rm vel}/(1-f_{\rm vel}) + \tau_{\rm S} f_{\rm ic}, 
 \label{Eq:fvel_fic}
\end{equation} 
and for the continuum case 
\begin{equation} 
 \chi_{\rm eff} \approx 1/h + \langle \chi \rangle f_{\rm
  ic}.  
\end{equation} 
These expressions show how the tenuous inter-clump medium now gradually fills in 
the porous channels (in velocity and physical space, for respectively lines and 
continuum) between the black clumps. 

As discussed in SPO14, while in the 
clump+void model the opacity \textit{itself} 
saturates (and thus becomes independent of the mean opacity), in this 
more general two-component model the \textit{ratio} between the effective 
and mean opacities (optical depths) saturates; thus the medium can 
always become effectively optically thick provided this 
mean opacity is high enough. 

The following sections now present some general, first results of our 
new clumpy numerical  {\sc fastwind} models.     
As directly evident from above, the key effect of 
optically thick clumps in the wind regards a potential reduction 
of opacities \textit{as compared to a wind consisting of only 
optically thin clumps.} This may then, in turn, affect both 
the ionization balance and the derivation of synthetic 
observables, where a key point is that the potential opacity 
reduction depends on the clump optical depth, and so 
becomes highly frequency and process dependent. 

\begin{table*}
        \centering
        \caption{Stellar, wind and clumping parameters for models in Sect. 4. For simplicity, all 
        models here assume clumping starts at $\varv/\varv_\infty=0.05$, increases linearly 
        until the value given in table below is reached at $\varv/\varv_\infty=0.1$, and then stays 
        constant throughout the rest of the wind. All models further assume a velocity-stretch porosity law 
        $h/h_\infty = \varv/\varv_\infty$ (see text) and solar metallicity. However, 
        for the $\zeta$ Pup like models the He and CNO abundances are 
        modified from the solar scale according to $n_{\rm He}/n_{\rm H}= 0.16$, and $A_{\rm C} = 6.73$, 
        $A_{\rm N} = 8.7$, $A_{\rm O} = 8.48$ on the standard abundance scale 
        $A_{\rm X} = \log^{10} (n_{\rm x}/n_{\rm H}) +12$.} 
                \begin{tabular}{p{2.8cm}lllllllllll}
                \hline \hline Name & $T_{\rm eff} \, \rm [K]$ & $\log g$ [cgs] & $R_\ast/R_\odot$ & 
                $\log \dot{M} \, \rm [M_\odot/yr]$ & $\varv_\infty \, \rm [km/s]$ & $\beta$ & 
                $L_{\rm x}/L_{\rm Bol}$ & $f_{\rm cl}$ & 
                $f_{\rm ic}$ & $f_{\rm vel}$ & $h_\infty/R_\ast$ \\ 
                \hline \\ 
                O1 smooth & 40\,000 & 3.6 & 19.0 & -5.22 & 2\,200 & 1.0 & 
                0 & 1.0 & - & - & - \\ 
                  O1 thin & 40\,000 & 3.6 & 19.0 & -5.22 & 2\,200 & 1.0 & 
                0 & 10.0 & - & - & - \\ 
                O1 thick & 40\,000 & 3.6 & 19.0 & -5.22 & 2\,200 & 1.0 & 
                0 & 10.0 & 0.01 & 0.5 & 1.0 \\ 

           	   O2 smooth & 35\,000 & 3.45 & 20.0 & -5.22 & 2\,000 & 1.0 & 
                0 & 1.0 & - & - & - \\ 
 		 O2 thin & 35\,000 & 3.45 & 20.0 & -5.22 & 2\,000 & 1.0 & 
                0 & 10.0 & - & - & - \\ 
  		O2 thick & 35\,000 & 3.45 & 20.0 & -5.22 & 2\,000 & 1.0 & 
                0 & 10.0 & 0.01 & 0.5 & 1.0 \\
                
                O3 smooth & 30\,000 & 3.2 & 22.0 & -5.22 & 1\,600 & 1.0 & 
                0 & 1.0 & - & - & - \\ 
                     O3 thin & 30\,000 & 3.2 & 22.0 & -5.22 & 1\,600 & 1.0 & 
                0 & 10.0 & - & - & - \\ 
                    O3 thick & 30\,000 & 3.2 & 22.0 & -5.22 & 1\,600 & 1.0 & 
                0 & 10.0 & 0.01 & 0.5 & 1.0 \\ 
                
                   B7 smooth & 12\,000 & 1.8 & 100.0 & -5.52 & 400 & 1.0 & 
                0 & 1.0 & - & - & - \\ 
              B7 thin & 12\,000 & 1.8 & 100.0 & -5.52 & 400 & 1.0 & 
                0 & 10.0 & - & - & - \\ 
                B7 thick & 12\,000 & 1.8 & 100.0 & -5.52 & 400 & 1.0 & 
                0 & 10.0 & 0.01 & 0.5 & 1.0 \\ 
                
              % $\sim$\,Zeta Pup smooth & 40\,000 & 3.63 & 18.9 & -5.74 & 2\,250 
              %  & 0.9 & $\sim10^{-7}$ & 1.0 & - & - & - \\ 
                  $\sim$\,$\zeta$ Pup thin & 40\,000 & 3.63 & 18.9 & -5.74 & 2\,250 
                & 0.9 & $\sim10^{-7}$ & 20.0 & - & - & - \\ 
                  $\sim$\,$\zeta$ Pup thick1 & 40\,000 & 3.63 & 18.9 & -5.74 & 2\,250 
                & 0.9 & $\sim10^{-7}$ & 20.0 & 0.01 & 0.5 & 1.0 \\ 
                $\sim$\,$\zeta$ Pup thick2 & 40\,000 & 3.63 & 18.9 & -5.74 & 2\,250 
                & 0.9 & $\sim10^{-7}$ & 20.0 & 0 & 0.5 & 1.0 \\               
                 
                \hline
                \end{tabular}
        \label{Tab:params}
\end{table*}

\subsection{Effects on ionization and opacities}

%Naturally, we have carefully tested so that our numerical implementation 
%recovers all basic limits discussed in the previous section (smooth 
%limit, optically thin, etc). With such basic verification in hand, 

%We focus now 
%in particular on verifying some of the limits derived 
%analytically in \S3.3.   

\paragraph{Ionization balance.} 

For the O-star models given in Table 1, Fig. \ref{Fig:ion} plots 
the ionization balance of (here the background elements) iron, 
phosphorus, and carbon, 
comparing now {\sc fastwind} simulations accounting fully 
for clumping of arbitrary optical thickness (blue lines), 
with models assuming optically thin clumps (red) and a 
smooth wind (black). The clumping parameters have 
characteristic values chosen from current theoretical 
and observational constraints: $f_{\rm cl} = 10$ and $h_\infty = R_\star$ 
are typical values found both in LDI models and observational 
studies \citep[e.g.,][]{Puls06, Leutenegger13, Sundqvist13, Grinberg15} 
and $f_{\rm ic} = 0.01$ and $f_{\rm vel} = 0.5$ are consistent with findings 
from the LDI \citep[see][]{Sundqvist10}. Regarding the velocity filling 
factor, we note that since $f_{\rm vol} \approx 1/f_{\rm cl} = 1/10$, 
$f_{\rm vel} = 0.5$ represents a rather high 
$| \delta \varv/\delta \varv_{\rm sm} | \approx 10$; such values for 
the clump velocity spans are indeed what is typically found in present-day 
LDI simulations and models trying to mimic these (\citealt{Sundqvist10}, 
e.g., see their Table 2).       

Overall, Fig. \ref{Fig:ion} demonstrates clearly that 
quantitative effects vary significantly between elements 
and stellar parameters; nonetheless, it illustrates also a few quite 
general effects of clumping upon the wind ionization balance. 
In particular: {\it i)} the increased rates of recombination in models with 
optically thin clumps generally imply a lower degree of ionization 
than in corresponding smooth simulations; {\it ii)} this effect  
is then somewhat counteracted by the light-leakage and effective 
opacity reduction associated with optically thick clumps. 

This quite general behavior can be understood by considering 
a simplified situation accounting only for ionization/recombination 
from/to the ground state of ion stage $j$ to/from the ground state of 
stage $j+1$. Following, e.g., \citet{Puls05}, an approximate NLTE number density 
ionization equilibrium equation can then be 
written\footnote{assuming a frequency dependence
of ionization cross-sections of the form $(\nu_{\rm edge}/\nu)^2$, which
is typical for metals.}  as (their eqn. 18):  
\begin{equation} 
	n_{j} \approx n_{j+1} n_{\rm e} \frac{\Phi(T_{\rm e})}{W(r)} \frac{T_{\rm e}}{T_{\rm rad,j}}
	\exp{ (\frac{h \nu}{k_{\rm b}} (\frac{1}{T_{\rm rad,j}}-\frac{1}{T_{\rm e}}) ) }, 
	\label{Eq:ion} 
\end{equation}
where $n_{\rm e}$ is the electron density, $\Phi(T_{\rm e})$ the 
Saha-Boltzmann factor (e.g., \citealt{Hubeny14}, their eqn. 9.5) for electron temperature 
$T_{\rm e}$, and the radiation temperature $T_{\rm rad}$ is defined from 
the mean intensity as $J(\nu,r) \equiv W(r) B_\nu(T_{\rm rad}(\nu,r))$ 
for Planck function $B_\nu$ and dilution factor 
$W(r) \equiv (1-\sqrt{1-R_\ast/r})/2$. 

For a major ion stage
$n_{j+1}/n_i \approx 1$, the product $n_{j+1} n_{\rm e} \sim \rho^2$ 
and so illustrates the first general effect discussed above (that clumped models 
tend to have lower degrees of ionization). But eqn. 
\ref{Eq:ion} also demonstrates explicitly the rather strong dependence on 
the local radiation field, where the exponential term on the right-hand-side 
decreases rapidly with an increasing $T_{\rm rad}$. For the three 
"O1" models in Table 1, Fig. \ref{Fig:trad} plots radiation temperatures
in a wavelength range around important ionization edges in the extreme 
ultra-violet, for a typical wind radius $r/R_\ast \approx 2$.  
We note from the figure that around and shortward of the 
second helium ionization edge at $228\, \AA$, the 
porous model yields radiation temperatures that are 
higher than in the simulation assuming optically thin clumps and more comparable to 
those seen in the smooth model (i.e., in this particular example 
$T_{\rm rad}^{\rm smooth} \approx T_{\rm rad}^{\rm thick} > T_{\rm rad}^{\rm thin}$). 
This is caused by the fact that 
$\chi_{\rm eff} = \langle \chi \rangle \ge \chi_{\rm smooth}$ in the model assuming 
optically thin clumps, whereas for the porous model 
$\chi_{\rm eff} \le \langle \chi \rangle$ (depending 
on $\tau_{\rm cl}$ at the relevant frequency and radius, see also Fig. \ref{Fig:kappa_eff}). 
In other words, porous models have reduced opacities when \textit{compared to 
equivalent models assuming optically thin clumping}, which is an upper 
limit. And while changes may appear quite modest at first glance, 
we recall from eqn. \ref{Eq:ion} that increasing $T_{\rm rad}$ by $\sim 1000$\,K in 
this range around $\sim 30\,000$\,K already reduces $n_j$ 
by more than a factor of two (assuming $n_{j+1} n_{\rm e}$ and $T_{\rm e}$ 
remain unaltered).     

Indeed, we can see this general ionization effect from a modified $T_{\rm rad}$ 
directly in our clumpy {\sc fastwind} models, for example 
in the ionization of trace ions of carbon in the outer 
wind: the upper right panel of Fig. \ref{Fig:ion} shows 
that optically thin clumping (red dashed line) reduces the relative 
number of CV atoms in the wind (as compared to the smooth model, 
the black dashed line), but 
how the addition here of porosity in physical and velocity space (blue dashed line)
then increases the ionization to levels quite close to those of the
smooth comparison model; the same behavior 
is also visible in the cooler `O3' carbon model to the lower 
right, however now it is the relative ionization of the 
minority stage CIV that is affected. Moreover, we note also that effects 
on the PV ion balance are quite moderate in the displayed models, thus providing some 
support of earlier phosphorus studies neglecting these ion feedback effects of 
velocity-space porosity \citep{Oskinova07, Sundqvist10, Sundqvist11, Surlan13}. 
 
 \paragraph{Continuum + line background opacities.} 
For parameters of the $\zeta$ Pup like 
model 'thick1' in Table 1, 
Fig. \ref{Fig:kappa_eff} shows a contour-plot of the ratio 
effective to mean opacities. 
In order to illustrate both the effects from line blocking/blanketing and the 
continuum, we have summed up the contributions from background lines and \textit{all} 
continuum opacities (i.e. from both background and explicit 
elements) when constructing this plot. The figure
shows that, for this $\zeta$ Pup like model, the strong 
background opacities in the extreme ultraviolet are reduced 
to their saturation value $\chi_{\rm eff}/\langle \chi \rangle = 0.01$ set 
by the assumed inter-clump medium density (see eqn. 29), 
whereas no effects of optically thick clumps are visible in the 
optical waveband. This illustrates that although the $\zeta$ Pup 
wind is optically thin in the visual waveband, the much higher 
opacities in the extreme ultra-violet are still subject to 
strong (velocity-)porosity effects. 

Moreover, the figure illustrates 
that (the primarily bound-free) high-energy opacities below $\la 10-20 \AA$ are 
barely affected when terminal porosity-lenghts are 
on order $h_\infty \la R_\ast$; this is consistent with previous 
porosity-analysis of observed X-ray line-profiles in this 
high-energy regime \citep{Leutenegger13}. 

Finally, the plot demonstrates that at low
wind radii continuum opacities at long radio 
wavelengths also reach the saturation value $\chi_{\rm eff}/\langle \chi \rangle = 0.01$; 
however, $\langle \chi \rangle$ is not affected at all in the outer wind beyond 
$\sim 10 R_\ast$. And since the radio photosphere for O-supergiants typically 
is located well above 
$10 R_\ast$, observations of radio emission in O-stars may
thus be relatively free from severe porosity effects; this will be 
further investigated in an upcoming paper (del-Mar Rubio et al., in prep.). 

\subsection{Effects on line profiles}

\paragraph{Ultra-violet resonance lines.} For 
major-ion resonance lines with $\langle \chi \rangle \sim \rho$, 
optically thin clumping affects computed profiles 
only through the potentially changed ionization balance (see above).
On the other hand, because of their very strong and 
localized opacities, these lines are the ones most prone 
to experience effects of porosity in velocity space. 
%A simple 
%order-of-magnitude estimate of the clump optical depth in  
%a model with negligible inter-clump density 
%
%\begin{equation} 
%	te
%\end{equation} 
%
%More explicitly, 

Fig. \ref{Fig:uv} 
shows the classical PV and NV `P Cygni' line doublets 
for the three $\zeta$ Pup like models of Table 1.  
The modeled phosphorus lines (upper panel) illustrate clearly the 
de-saturating effect of velocity-space porosity: 
The blue line shows strong PV lines emerging 
from simulations assuming optically thin clumping; the 
red dashed line then illustrates how velocity-porosity leads 
to significant additional light escape and so to weaker line 
profiles for a given $\dot{M}$; finally, the black line 
shows how a tenuous inter-clump medium density of 
$f_{\rm ic} = 0.01$ does not significantly affect this line
(simply because the inter-clump densities are too low to 
contribute much to the total, effective line opacity). In agreement 
with previous studies \citep{Oskinova07, Sundqvist10, 
Sundqvist11, Surlan13, Sundqvist14}, this figure thus 
demonstrates explicitly how porosity in velocity space 
provides a natural explanation of the weak  
PV lines observed for O-stars in the Galaxy 
\citep{Fullerton06}.  

The lower panel then shows the same principal effects 
for the strong NV resonance doublet. Moreover, 
the red dashed line profile here also serves to demonstrate 
how a model with an effectively void medium actually 
never becomes completely `black', since the 
absorptive intensity saturation value 
$e^{-\tau_{\rm eff} (f_{\rm vel})}$ ($=e^{-1}$ for the $f_{\rm vel}=0.5$ 
assumed here) becomes independent of the atomic opacity
(see eqn. \ref{Eq:fvel} and previous discussion). However, unlike the weaker PV doublet, the strong 
NV lines are also sensitive to the inter-clump medium density; 
this is illustrated here by the black showing stronger absorption 
for the $f_{\rm ic} =0.01$ model than for the effectively 
void one (red dashed line). Physically,  
increasing the inter-clump density means 
the velocity-space holes between clumps are
gradually ``filled" in by absorbing material, 
until eventually the absorption profile 
becomes ``black'' at some saturation value 
$f_{\rm ic} \tau_{\rm S}$ (see eqn. \ref{Eq:fvel_fic}), which 
here clearly lies at $f_{\rm ic} > 0.01$.  
%As discussed further in \S5, 
This essentially demonstrates 
how a combination of 
saturated and un-saturated UV P-Cygni lines 
may be used to constrain properties of both the 
high-density clumps and the sparse wind 
plasma in between them.  

\paragraph{$H_\alpha$.} As discussed extensively 
in \citet{Sundqvist11} (see also \citealt{Oskinova07}), 
for O-stars the optical $H_{\alpha}$ line is generally 
less sensitive to velocity-space porosity effects than the 
UV resonance lines above. This is illustrated in the 
upper two panels of Fig.~\ref{Fig:ha}, displaying only 
small differences (actually quite negligible when 
the line is used for deriving $\dot{M}$)  
between the ``thin'' and ``thick'' clumping 
simulations of the $\zeta$ Pup like and 03 models 
in Table 1. The marginal increase in H$\alpha$ 
line emission of the '`thick'' models here is due to a 
small change in the NLTE number densities 
for the involved hydrogen atomic levels (neglected 
in previous work, which has only considered velocity-space 
porosity in the ``formal integral'' and not in the calculation of 
the NLTE number densities). 
Moreover, the principal recombination-line 
emission-measure scaling $\chi \sim \langle \rho^2 \rangle \sim \dot{M}^2 f_{\rm cl}$ 
(Sect. 4.1) is directly visible in the middle panel 
through the much stronger profiles of the clumped models than 
the smooth one. From the $\zeta$ Pup like models, we note finally that, as expected,
the inter-clump density is not very important for the 
H$\alpha$ line formation in this regime.  

The lowest panel in Fig.~\ref{Fig:ha} shows, however, that when changing 
the stellar parameters to those of typical late B- and 
A-supergiants, we see effects quite similar to those of the UV 
lines analyzed above. This was pointed out already by \citet{Sundqvist10}, 
but with our new {\sc fastwind} models we are now able to 
quantify this potentially important effect. Namely, in this regime 
the lower level of the H$\alpha$ transition becomes much more 
populated, and so the line transforms from being an almost optically thin 
recombination line to being much more optically thick
\citep[e.g.,][]{Kudritzki00}. As such, H$\alpha$ is 
much more prone to velocity-space porosity effects in this regime 
than for O-stars (as clearly seen in the lower panel of 
Fig.~\ref{Fig:ha}).  
  
%\paragraph{Temperature structure.} 

\section{Summary and future work} 

We have developed a formalism that accounts 
for the leakage of light associated with 
porosity in physical and velocity space in a clumpy, 
accelerating medium, and incorporated this 
method into the global (photosphere+wind) 
NLTE model atmosphere code {\sc fastwind}. 
Our method is included both in the 
general code network for solving the NLTE rate 
equations and in the ``formal solver" used to 
produce synthetic spectra. As such, the new version 
of {\sc fastwind} allows readily for investigations of optically 
thick clumping directly on line profiles, but also upon the wind 
ionization balance and NLTE number densities. Indeed, 
to date it is the only stellar atmosphere code on the market that includes
these physical effects of clumping in the NLTE network; along with 
its quite remarkable computational speed 
(see below), the new version of {\sc fastwind} is thus ideally suited for 
future multi-wavelength quantitative spectroscopic studies 
aiming to derive new improved wind parameters of individual 
stars in the local Universe. 
%(and also, e.g., for broader studies of 
%integrated UV light of unresolved populations of massive 
%stars in starburst galaxies at cosmological distances). 

The basic method is based on a description of the supersonic 
wind outflow as a stochastic two-component medium 
consisting of dense clumps (of arbitrary optical thickness) and a 
rarified `inter-clump' medium (filling in the space between the
clumps). By formulating this in terms of effective and mean opacities 
scaled to corresponding ``smooth'' ones, we are able to compute 
models with the same speed as in previous 
{\sc fastwind} programs ($\sim$\,15 minutes on a modern laptop/desktop). 
Four parameters are required to fully specify 
the clumpy two-component medium; the clumping 
factor $f_{\rm cl}$, the inter-clump density parameter $f_{\rm ic}$, 
the porosity length $h$ and the velocity filling factor $f_{\rm vel}$. 
While the first three of these parameters are the equivalents of those
used in previous stochastic transport models for static 
media \citep{Pomraning91}, the last one accounts here 
for line-absorption in accelerating, supersonic media (SPO14). 
As discussed in Sect. 3, we note also that the set 
$f_{\rm cl}$, $h$, and $f_{\rm vel}$ is equivalent to a 
consideration of the clump volume filling factor, the clump 
length scale, and the clump velocity-span. 

After confirming that relevant analytic limits are preserved (e.g., smooth, 
optically thin and completely optically thick cases), 
we present some first, generic {\sc fastwind} results. A summary is 
as follows: i) We confirm earlier results \citep[e.g.,][]{Oskinova07, Sundqvist11, 
Surlan13, Sundqvist14} that velocity-space porosity is critical for 
UV wind resonance lines in O-stars, and indeed provides a natural 
explanation for the overall weakness of, e.g., observed unsaturated 
PV lines in the Galaxy. In addition, we illustrate how a combination of 
saturated and un-saturated UV lines might be used to constrain not only the dense 
clumps, but also the rarefied material in between them (under the assumption made 
here of approximately equal gas and radiation temperatures for both components). 
ii) For the optical H$\alpha$ line, we show that optically thick clumping effects 
are marginal for O-stars, but potentially very important for late 
B and A-supergiants (see also below). iii) In agreement with previous 
work, we show that spatial porosity is a marginal effect for absorption of high-energy 
X-rays below, say, $\sim 15-20 \rm \AA$ in O-stars, as long as the terminal 
porosity lengths are kept at realistic values $h_\infty \la R_\ast$ 
\citep{Leutenegger13, Herve13, Grinberg15, Owocki18}. iv) Also, 
whereas radio absorption indeed shows strong spatial porosity effects in the 
inner wind, it is negligible at typical O-star radio-photosphere radii 
$r \sim 100 R_\ast$. v) Regarding the wind ionization balance, quantitative 
effects depend significantly on the specific ion and stellar and wind 
parameter range. However, two general trends are 1) the increased rates of 
recombination in simulations with optically thin clumps lead to overall lower 
degrees of ionization than in corresponding smooth models and 2) this effect 
is then counteracted by the increased light-leakage
(i.e., the stronger radiation field) associated with porosity (in physical and/or 
velocity space). 

An overall objective for the new version of {\sc fastwind} now concerns 
application to multi-wavelength observations of samples of hot, massive 
stars, and corresponding derivation of empirical constraints on clumping 
parameters and mass-loss rates. Indeed, the paper here shows that, if enough 
suitable diagnostics are available one should be able to obtain direct 
constraints on the different clumping parameters (and their radial dependence). 
A rough outline for a 4-step procedure of such multi-wavelength analyses could 
be as follows: i) Use the relatively clumping-insensitive X-ray lines and/or broad-band 
spectrum to derive an empirical mass-loss rate \citep{Cohen11, Herve13, Cohen14}; ii) then, 
obtain clumping factors for the near-photospheric layers using optical and/or infra-red (IR)
recombination lines like H$\alpha$, HeII 4686, Br$\alpha$, etc \citep{Puls06, Sundqvist11, Najarro11};
iii) then, obtain constraints on clumping factors in the intermediate/outer wind from IR/radio 
emission \citep{Puls06} and (at least) upper limits for porosity lengths by combining 
IR/radio and X-ray \citep{Leutenegger13} results; iv) then, obtain constraints on the velocity filling 
factor and the inter-clump density from a combination of saturated and un-saturated UV line 
profiles from different ions (see discussion above). 
%; v) finally then, go back and check so your results are 
%consistent. Iterate if necessary.      
      
However, while this overall outline provides neat guidance for carrying out 
extensive analyses of nearby stars with a multitude of observational material 
available, for other cases (e.g., single line observations) one will have to rely on
external constraints and knowledge for setting the clumping parameters.  For such cases, reasonable input 
values (in overall agreement with various theoretical and empirical constraints, see also \S4.2) might be 
$f_{\rm ic} \sim 0.01-0.2$, $f_{\rm vel} \sim 0.5-1.0$, $h_\infty/R_\ast \sim 0.5-1.0$, and $f_{\rm cl} \sim 4 -30$. 
Also, let us repeat here again that, in practice, certainly not all clumping 
parameters will be relevant for all diagnostics (see previous sections). 
 
In addition to UV P-Cygni lines in OB-stars, one interesting 
application regards H$\alpha$ in late B and A-supergiants, 
for which the observed absorption troughs very often are shallower than 
predicted by smooth/optically thin clumping models (M. Urbaneja, private 
communication); this agrees well with the generic results from models including 
velocity-space porosity (Fig. \ref{Fig:ha}, lower panel), and definitely deserves further studies. 
Indeed, such velocity-porosity effects may potentially be important also 
for quantifying mass loss from H$\alpha$ across the broader B-star regime \citep{Petrov14}, 
in particular in regions where the iron ion balance undergoes a quite sudden 
shift (at the so-called bistability-jump, \citealt{Pauldrach90, Vink00}); this will 
be analyzed in detail in a forthcoming paper (Watts et al. in prep.). And in yet 
another waveband, a detailed analysis of (the potential lack of) porosity effects for 
deriving radio mass-loss rates from O-stars is also underway (del-Mar Rubio 
et al., in prep., but see also \citealt{Ignace16} for alternative models). 

Finally, key steps in future updates of {\sc fastwind} will regard i) a CMF-solution of ``all'' 
contributing lines \citep[see][]{Puls17}; ii) an extension toward modeling of optically 
thick Wolf-Rayet winds (where spatial/velocity-field porosity effects definitely should be very important, 
Sundqvist et al., in prep); and iii) solving the (steady-state) hydrodynamical equations 
in the wind in order to also predict average mass-loss rates and velocity fields (Sundqvist et al., in 
prep, but see also \citealt{Sander17, Krticka17} for alternative models).  

\begin{acknowledgements}
  JOS acknowledges previous support from DFG grant Pu117/8-1, 
  and also previous support from the European UnionÕs Horizon 
  2020 researchand innovation program under the Marie 
  Sklodowska-Curie grant agreement No 656725. JP gratefully acknowledges 
  travel support by the Deutsche Forschungsgemeinschaft, under grant Pu117/9-1.
  We finally thank the referee, Ken Gayley, for useful comments on the manuscript. 
\end{acknowledgements}

%\citestyle{aa}
\bibliographystyle{aa}
\bibliography{fastwind4}

\end{document}